\documentclass[pra,amsmath,amssymb, aps,graphicx,onecolumn,reprint]{revtex4-2}
\usepackage{graphicx}
\usepackage{physics}
\usepackage{color}
\usepackage{mathrsfs}
\usepackage{threeparttable}

\begin{document}
\newcommand{\dif}{\mathrm{d}}

\title{Experimental preparation of generalized cat states for itinerant microwave photons}

\author{Zenghui Bao}
\altaffiliation{These two authors contributed equally to this work.}

\author{Zhiling Wang}
\altaffiliation{These two authors contributed equally to this work.}

\author{Yukai Wu}

\author{Yan Li}

\author{Weizhou Cai}

\author{Weiting Wang}

\author{Yuwei Ma}

\author{Tianqi Cai}

\author{Xiyue Han}

\author{Jiahui Wang}

\author{Yipu Song}

\author{Luyan Sun}

\author{Hongyi Zhang}
\email{hyzhang2016@tsinghua.edu.cn}

\author{Luming Duan}
\email{lmduan@tsinghua.edu.cn}

\affiliation{Center for Quantum Information, Institute for Interdisciplinary Information Sciences, Tsinghua University, Beijing 100084, PR China}

\date{\today}

\begin{abstract}

Generalized cat states represent arbitrary superpositions of coherent states, which are of great importance in various quantum information processing protocols. Here we demonstrate a versatile approach to creating generalized itinerant cat states in the microwave domain, by reflecting coherent state photons from a microwave cavity containing a superconducting qubit. We show that, with a coherent control of the qubit state, a full control over the coherent state superposition can be realized. The prepared cat states are verified through quantum state tomography of the qubit state dependent reflection photon field. We further quantify quantum coherence in the prepared cat states based on the resource theory, revealing a good experimental control on the coherent state superpositions. The photon number statistic and the squeezing properties are also analyzed. Remarkably, fourth-order squeezing is observed in the experimental states. Those results open up new possibilities of applying generalized cat states for the purpose of quantum information processing.

\end{abstract}

\maketitle 

\section{Introduction}

Coherent states are regarded as quasi-classical states with minimum uncertainties. Preparing superpositions of coherent state is of widespread interest to explore such an intriguing quantum phenomenon for macroscopically distinct states~\cite{Wineland2013,Haroche2013,Polzik2006,Ourjoumtsev2006,Ourjoumtsev2007,Takahashi2008,Grangier2009,Knill2010,Namekata2010,Schoelkopf2013,wang2016,Alexander2017,Schoelkopf17,Rempe2019,Wallraff2020}. Specifically, equal superpositions of coherent states $\ket{\alpha} \pm \ket{-\alpha}$ are known as cat states, in deference to Schrödinger's famous thought experiment~\cite{Sch1935}. Accordingly, arbitrary superpositions of coherent states are commonly termed as generalized cat states~\cite{Rempe2019}, which serve as important quantum resources for various continuous-variable based quantum information processing protocols, including quantum communication~\cite{vanLoock2008,Jiang2017prl}, quantum computation~\cite{Munro1999,Kim2002,Ralph2003,Ralph2008} and quantum metrology~\cite{Gilchrist_2004,Sciarrino2020}. 
Even though the cat states are known to be delicate, they have been demonstrated as a powerful encoding scheme for the implementation of quantum error correction~\cite{Leghtas2013,Mirrahimi2014,Ofek2016}. 

Schrödinger’s cat is not necessarily confined in a closed box, but could take a propagating mode. Itinerant cat states of photons have been proposed to be used as carriers of quantum information through a lossy channel~\cite{Jiang2017prl,Jiang2017}, or for fault-tolerant quantum computation~\cite{su2021universal,hastrup2021alloptical}. Generalized itinerant cat states are originally prepared with linear optics in a nondeterministic approach by subtracting single photon from a squeezed vacuum and coherent displacement operations~\cite{Sasaki2010}. A deterministic preparation requires a cavity quantum electrodynamics (QED) system, with which odd and even cat states have been successfully prepared by either releasing a stationary cat from the cavity as propagating modes~\cite{Schoelkopf17} or reflecting coherent state photons from a cavity containing a qubit~\cite{Wallraff2020,wang2021flying}. The latter scheme is essentially a variation of Duan-Kimble scheme for optical quantum computation~\cite{Duan04cz}, which has been broadly used in quantum non-demolition measurement of a single photon~\cite{Duan04cz,Rempe2013,Nakamura2018,Wallraff2018,Wallraff2020}, realizing controlled-phase gate between an atomic and a photonic quantum bit~\cite{Rempe2014,Lukin2014} or between two photonic quantum bits~\cite{Duan04cz,Rempe2016}, and generating remote qubit entanglement~\cite{Duan05remote,Rempe2021}. Notably, generalized cat states in the optical domain have been successfully prepared with this scheme~\cite{Rempe2019}.

In this work, we have prepared generalized itinerant cat states in the microwave domain with the above mentioned scheme.
Coherent state photons are reflected from a microwave cavity containing a superconducting qubit. A full control over the coherent state superposition can be realized by a coherent control of the qubit state. Quantum superpositions are verified with quantum state tomography and further quantified with a quantum coherence resource theory. The prepared states exhibit distinct non-classical features in the statistical properties of photon field, demonstrating a well-controlled evolution of the reflected photon field from a super-Poissonian distribution to a sub-Poissonian distribution, or from anti-squeezing to squeezing. In particular, fourth-order squeezing is observed in the superposition states. Those results demonstrate an important toolbox for quantum information processing protocols based on the cat states.

\begin{figure}[!tbp]
\centering
\includegraphics[width=1\linewidth]{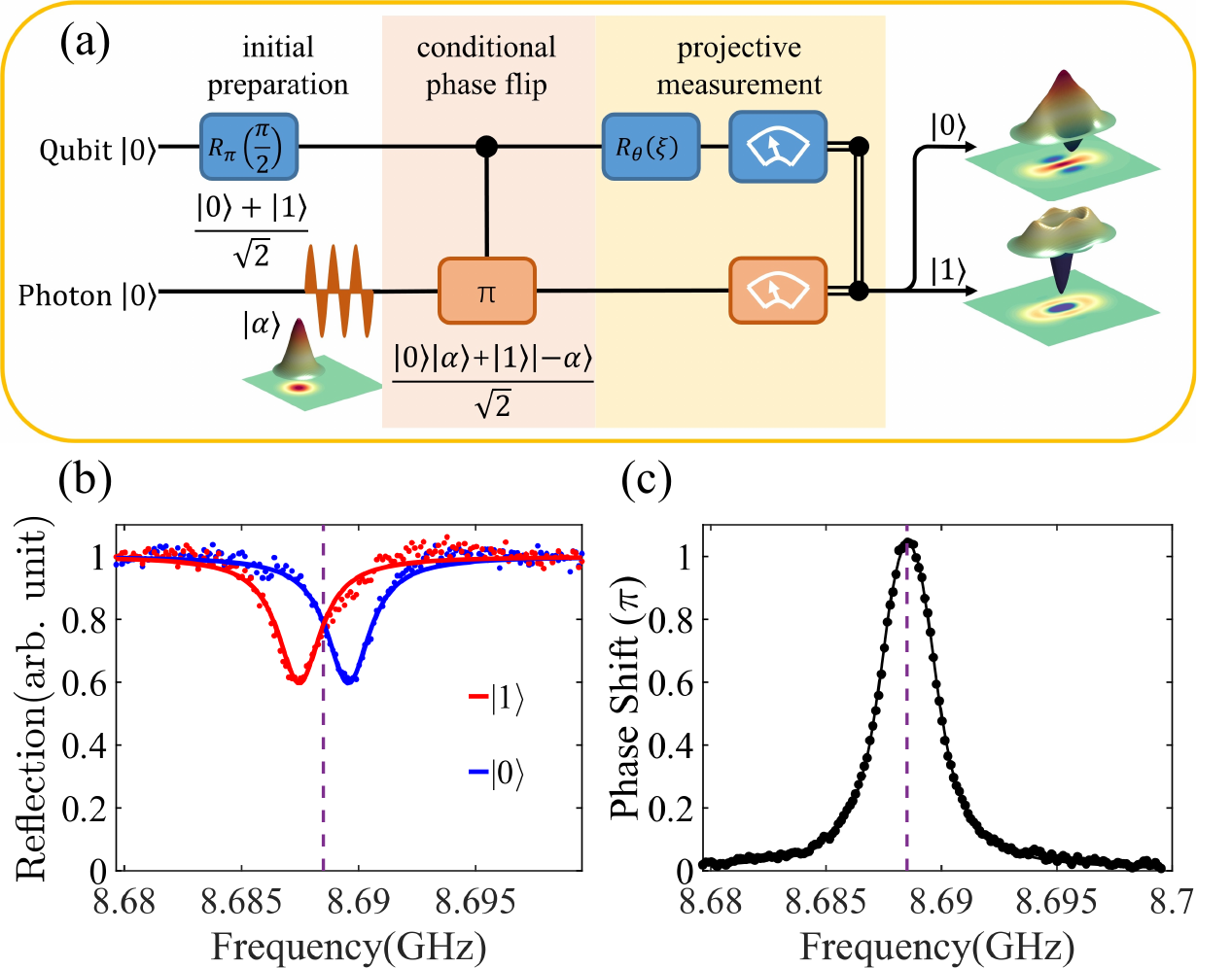}
\caption{\textbf{The protocol for the generalized cat state preparation.} (a) The pulse sequence for the preparation of generalize cat states. Initially, the qubit is prepared to a superposition state $(\ket{0}+\ket{1})/\sqrt{2}$, and coherent state microwave photons are sent to the cavity. The reflected photons would acquire a conditional phase depending on the qubit state. 
The qubit is further rotated with $R_{\theta_q}(\xi)=\exp(-i\frac{\xi}{2}(\sigma_x\sin\theta_q-\sigma_y\cos\theta_q))$, and projected to either $\ket{0}$ or $\ket{1}$ with a projective measurement. The reflected photons conditioned on the qubit state either in $\ket{0}$ or $\ket{1}$ would be a superposition of coherent states. 
(b) The measured cavity reflection spectra when the qubit is in either $\ket{0}$ (blue) or $\ket{1}$ (red). In experiment we tune the cavity linewidth to meet $\kappa \sim 2|\chi|$. The resulting (c) phase difference of the qubit-state-dependent reflections reaches $\pi$ at the bare cavity frequency $\omega_c$. The dots are measured results and the solid lines are theoretical fitting results. The dashed line indicates the frequency of the input photons.}
\label{illustration}
\end{figure}

\section{Methods}

We use a 3D microwave cavity dispersively coupled to a superconducting transmon qubit, for which the Hamiltonian can be written as $H/\hbar=(\omega_c+\chi\sigma_z)a^\dagger a+\omega_q\sigma_z/2$, where $a$ ($a^\dagger$) is the annihilation (creation) operator of the cavity mode, $\sigma_z$ is Pauli operator of the qubit, $\omega_c$ represents the cavity frequency, $\omega_q$ is the qubit frequency, and $\chi$ is dispersive shift induced by the interaction between the qubit and the cavity.
As shown in Fig.~\ref{illustration}(b), the cavity reflectance depends on the state of the qubit due to the dispersive term, resulting in a phase difference of the cavity reflectance when the qubit is in $\ket{0}$ and in $\ket{1}$. In the experiment we tune the total cavity line width $\kappa_{tot}$ to approximately $2|\chi|$ to realize a phase difference of $\pi$ at $\omega_c$, as illustrated in Fig.~\ref{illustration}(c). 

The protocol for the generation of generalized itinerant cat states in the microwave domain is shown in Fig.~\ref{illustration}(a). The qubit initially takes a quantum superposition $(\ket{0}+\ket{1})/\sqrt{2}$. When a coherent state microwave pulse is sent to the cavity, the reflected photons would acquire a conditional phase shift depending on the qubit state, and thus the system would be in a Schrödinger cat state expressed as~\cite{Duan05cat}
\begin{equation}
(\ket{0}\ket{\alpha}+\ket{1}\ket{-\alpha})/\sqrt{2}
\label{qph-entan}
\end{equation}
Such a photon-qubit entangled state serves as the basis for the generation of generalized cat states. 
If an arbitrary rotation $R_{\theta_q}(\xi)=\exp(-i\frac{\xi}{2}(\sigma_x\sin\theta_q-\sigma_y\cos\theta_q))$ is applied on the qubit, and then the qubit state is projected to either $\ket{0}$ or $\ket{1}$ with a projective measurement, we would have qubit state dependent photon state as 
\begin{equation}
\begin{split}
\ket{\psi_0}&=\mathscr{N}(\cos\frac{\xi}{2}\ket{\alpha}+\sin\frac{\xi}{2}e^{-i\theta_q}\ket{-\alpha})\\
\ket{\psi_1}&=\mathscr{N}(-\sin\frac{\xi}{2}e^{i\theta_q}\ket{\alpha}+\cos\frac{\xi}{2}\ket{-\alpha})
\end{split}
\label{general_cat}
\end{equation}
where $\mathscr{N}$ represents a suitable normalization factor. It can be seen that $\xi$ and $\theta_q$ control the weight and phase of the coherent state superposition, respectively.
Specifically, with $\xi = \pi/2$ and $\theta_q = 0$ we would have odd or even cat state of microwave photons $\ket{\psi_{e/o}}=\mathscr{N}(\ket{\alpha}\pm\ket{-\alpha})$ conditioned on the qubit state of either $\ket{0}$ or $\ket{1}$. If taking $\xi = \pi/2$ and $\theta_q = \pi/2$ we would have the so-called Yurke-Stoler (YS) state~\cite{Stoler1986} expressed as $\ket{\psi_{YS}}=\mathscr{N}(\ket{\alpha}\pm e^{-i\pi/2}\ket{-\alpha})$. The YS state differs from the odd or even cat state only by a superposition phase, but shows distinctly different photon number statistics, as discussed later. 

Before being acquired with a homodyne setup, the reflected photons are successively amplified by a cascade amplifier circuit containing a Josephson junction parametric amplifier (JPA), a high-electron-mobility-transistor amplifier and two microwave amplifiers at the room temperature. The JPA is working in a phase preserving mode~\cite{JPA1,JPA2} with a gain of 16 dB around the cavity frequency, yielding an overall circuit detection efficiency $1/(n_{noise}+1)=20\%$ with a noise photon number $n_{noise}=4$. In this way we record the quadrature distribution of the qubit state dependent reflection signal, and calculate its moments up to the sixth order. The quantum state of the reflected photons can be reconstructed from the moments with a maximum likelihood method~\cite{Eichler2011}. It is worth noting that statistical features of the photon field, such as photon number distribution and squeezing, can be directly extracted from the calculated moments~\cite{Eichler2011}.

\section{Results}

\subsection{Generalized cat states.} 

\begin{figure}[!tbp]
\centering
\includegraphics[width=1\linewidth]{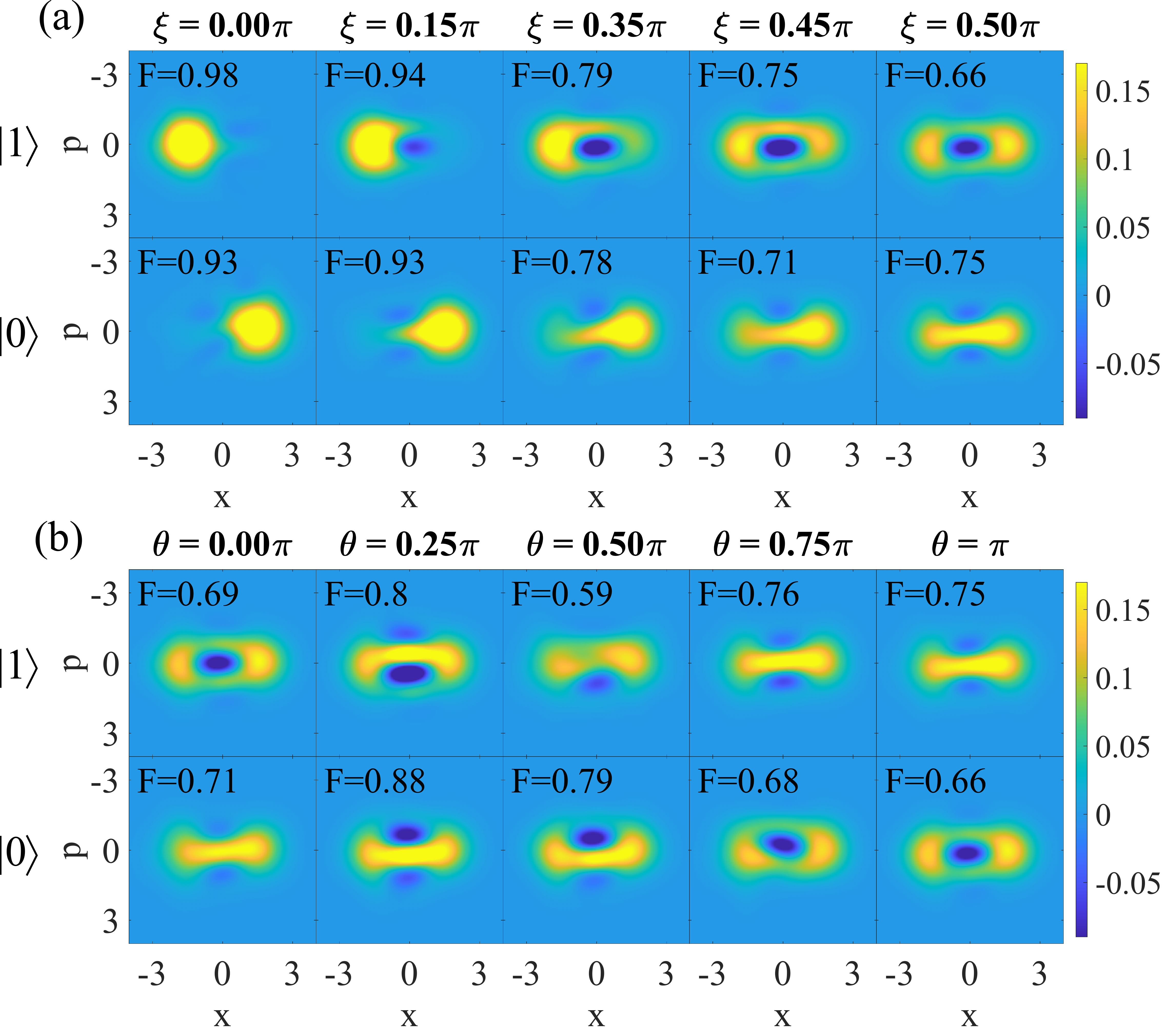}
\caption{\textbf{Experimental Wigner function of the generalized cat states.} For varied rotation angles $\xi$ and $\theta$ of the coherent state superposition, the measured Wigner functions of the reflected itinerant microwave photons conditioned on the qubit state in either $\ket{0}$ or $\ket{1}$. We choose the coherent state amplitude $\alpha=1.07$. (a) shows the results when scanning $\xi$ from 0 to $\pi/2$ while keeping $\theta = 0$. The reflected photon state gradually evolves from a coherent state to an equal superposition of two coherent states with opposite amplitudes, known as even or odd cat state when the qubit state is in $\ket{0}$ or $\ket{1}$. (b) shows the results when varying $\theta$ from 0 to $\pi$ while keeping $\xi = \pi/2$. The phase evolution of the cat state can be seen from the change of the interference fringes. Conditioned on the qubit state in $\ket{0}$ (or $\ket{1}$), the measured odd (even) cat state continuously evolves to an even (odd) cat state. The fidelities of the experimentally reconstructed state compared with the corresponding ideal state are labeled in each figure.}
\label{Generalizedcat}
\end{figure}

Following the protocol mentioned above, we experimentally showcase the preparation of arbitrary superpositions of two coherent states with opposite amplitudes. It is important to note that in the presence of finite cavity loss, the coherent state superposition phase $\theta$ deviates from the qubit superposition phase $\theta_q$, with $\theta - \theta_q = 0.125\pi$ corresponding to $\alpha=1.07$ in the experiment. Detailed discussions can be found in Appendix ~\ref{sec:app-D}. In Fig.~\ref{Generalizedcat}(a) and (b) we present the reconstructed Wigner functions of the reflected itinerant microwave photons conditioned on the qubit state in $\ket{0}$ and $\ket{1}$, with varied azimuthal angles $\theta$ and polar angles $\xi$ of the coherent state superposition. The negative value regions in the Wigner functions characterize quantum coherence for the prepared states, which are of no classical explanation.
By scanning $\xi$ from 0 to $\pi/2$ and using $\theta=0$, the reflected photon state changes continuously from a coherent state to cat states with equivalent distributions on $\ket{\alpha}$ and $\ket{-\alpha}$, as expected from Eq.~\ref{general_cat}.
When varying $\theta$ from 0 to $\pi$ while keeping $\xi=\pi/2$, the measured photon state gradually evolves from an even cat state to an odd cat state conditioned on the qubit state in $\ket{0}$, or vice versa for qubit state in $\ket{1}$, indicating a well-controlled superposition phase. Specifically, with $\theta=\pi/2$ we obtain the YS states, characterized by the unsymmetrical interference patterns~\cite{Stoler1986,Knight1992}. It is known that quantum interference of the coherent components alters the photon statistics, and thus conceivably results in different parities of the superposition state. In the experiment, we observe a continuous evolution from a sub-Poissonian photon distribution to a super-Poissonian one with a varied $\theta$ from 0 to $\pi$, which agrees well with the theory and manifests a well control over the photon number statistics of the superposition states. More details can be found in Appendix ~\ref{sec:app-E-3}. Those results unambiguously manifest the preparation of arbitrary quantum superposition of coherent states.

The size of the superposition state can be controlled by the amplitude of the coherent components $\alpha$. Superposition of coherent states with arbitrary optical phases can be also prepared with the current scheme. The optical phases, which are defined as the phase difference between two coherent components, are determined by the signal frequency, which in principle can be changed from $\pi$ to 0 through a detuning from the bare cavity resonance $\omega_c$. We showcase the preparation of superposition containing two coherent states with a relative optical phase of $0.9\pi$ with a detuning of $0.7$ MHz. Details of those results can be found in Appendix ~\ref{sec:app-E-2}.

The infidelity of the prepared cat states mainly originates from cavity loss during the reflection process, qubit decay/dephasing and qubit state measurement error. As for our experiment, cavity loss is the dominant error source for most of the cases, which contributes more than 60\% of infidelity to the experimentally prepared odd and even cat states with $\alpha=1.07$. Considering that the cavity internal loss rate $\kappa_i/2\pi$ is about 0.22 MHz in the experiment, which is substantially larger than the state-of-the-art values~\cite{Reagor2016}, an improved microwave cavity with smaller $\kappa_i$ is preferred to achieve better fidelity, especially for cat states with large $\alpha$. Detailed discussion about the error model and error budget can be found in Appendix ~\ref{sec:app-D}.

\begin{figure}[!tbp]
\centering
\includegraphics[width=1\linewidth]{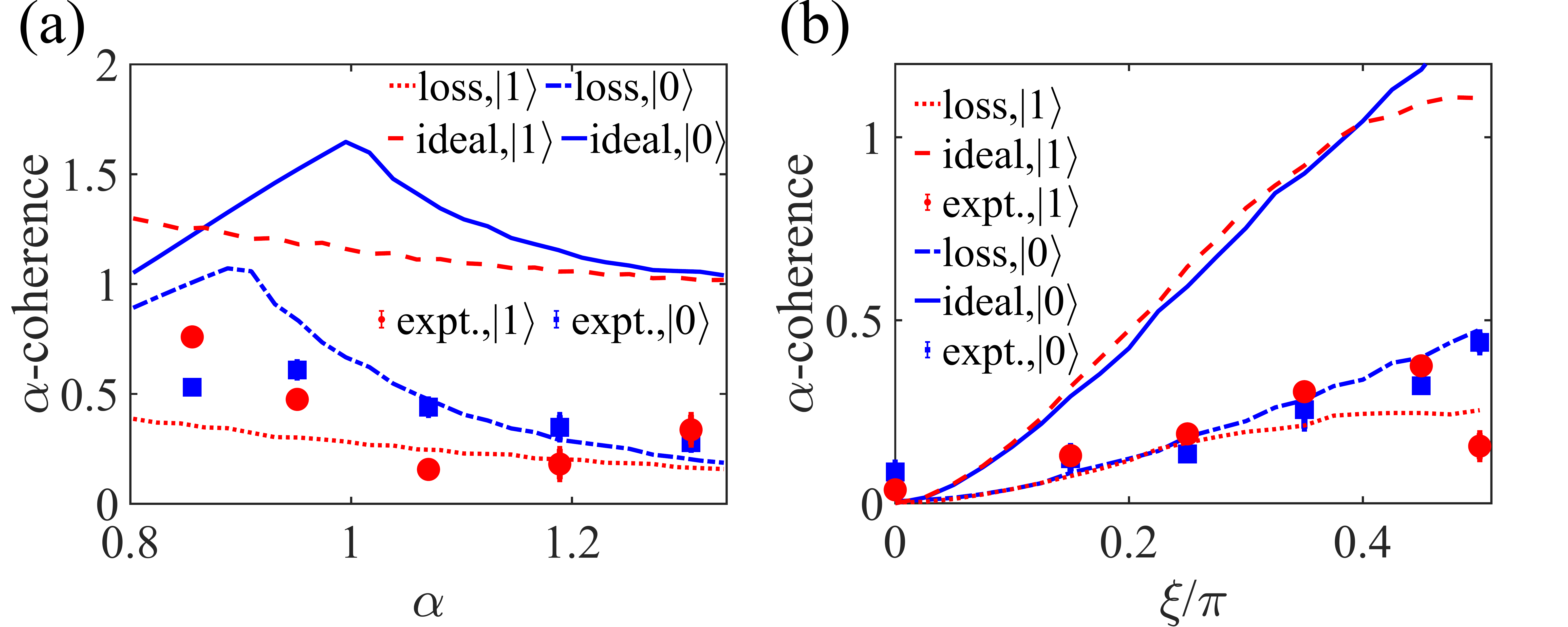}
\caption{\textbf{Quantum coherence and photon statistics of the generalized cat states.} (a) $\alpha$-coherence for the odd and even cat states with varied sizes. (b) $\alpha$-coherence of the superposition states conditioned on the qubit state in $\ket{0}$ and $\ket{1}$, when varying $\xi$ from 0 to $\pi/2$ while keeping $\theta$=0. The error bar is given by a standard deviation of the measured data, some of which are smaller than the size of the markers. The blue solid lines and the red dashed lines are theoretical results based on the corresponding ideal states conditioned on the qubit state in $\ket{0}$ and $\ket{1}$, respectively. The blue dotted dashed lines and the red dotted lines are theoretical results considering possible experimental loss and decoherence conditioned on the qubit state in $\ket{0}$ and $\ket{1}$, respectively. The blue squares and the red circles show the experimental data with for the states conditioned on the qubit state in $\ket{0}$ and $\ket{1}$, respectively.}
\label{Poissonian}
\end{figure}

\subsection{Quantum coherence of the cat states.}
To quantify the quantum superposition in the prepared states, we analyze the experimental states with a recently developed quantum coherence resource theory~\cite{Plenio2014,Jeong2017}. The resource theory essentially measures the amount of coherence introduced by superposition or entanglement in a give quantum state. As for coherent state based superposition, the idea of $\alpha$-coherence was introduced based on the Glauber-Sudarshan P distribution, assisted with proper orthonormalization of the basis set and state decomposition~\cite{Jeong2017}. In this definition, coherent states or their statistical mixtures are taken as classical states without coherence, correspondingly one would have $\alpha$-coherence as 0.

\begin{figure}[!tbp]
\centering
\includegraphics[width=0.9\linewidth]{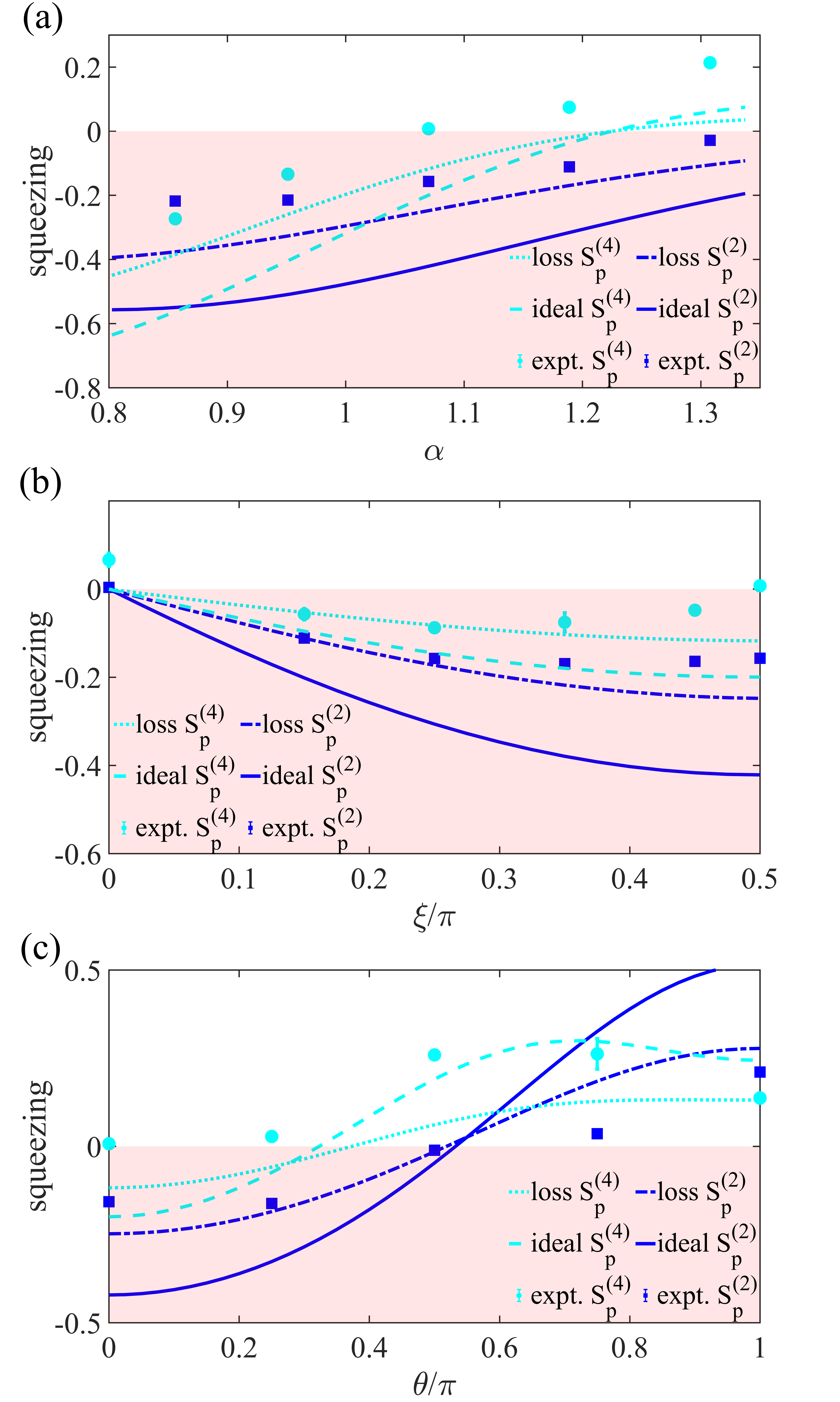}
\caption{\textbf{Squeezing of the generalized cat states.} The second-order and fourth-order squeezing for the various coherent state superpositions conditioned on the qubit state in $\ket{0}$, with (a) varied coherent state amplitude $\alpha$ when keeping $\xi=\pi/2$ and $\theta=0$, (b) varied population fraction of the two coherent components $\xi$ when using $\alpha=1.07$ and $\theta = 0$ or (c) varied superposition phase $\theta$ when using $\alpha=1.07$ and $\xi=\pi/2$. $S_p^{(N)}<0$ indicates the existence of squeezing. The scattered plots show the experimental data. The error bar is given by a standard deviation of the measured data, some of which are smaller than the size of the markers. The blue dotted dashed lines and cyan dotted lines are theoretical results of second-order and fourth-order squeezing considering possible experimental loss and decoherence, respectively. The blue solid lines and cyan dashed lines are theoretical results of second-order and fourth-order squeezing of the corresponding ideal states, respectively. The blue squares and the cyan circles show the experimental results of the second-order and fourth-order squeezing, respectively.}
\label{quantum_coherence}
\end{figure}

In Fig.~\ref{Poissonian}(a) and (b) we present $\alpha$-coherence for the experimental states with varied $\alpha$ and $\xi$. The non-zero values of $\alpha$-coherence characterizes the non-classical property of the prepared states.
When the prepared state evolves from a coherent state to an odd or even cat state by varying $\xi$ from 0 to $\pi/2$, one could see a continual growth of $\alpha$-coherence as shown in Fig.~\ref{Poissonian}(b). Theoretical results based on the error model are basically in line with the trend of experimental data, which verifies that the deviations between experimental states and their corresponding ideal states are mainly caused by cavity loss and qubit-related imperfections induced decoherence of the superposition states. This decoherence effect can also be reflected from the decreasing trend of $\alpha$-coherence of odd and even cat states with larger size, as shown in Fig.~\ref{Poissonian}(a). This is because the coherence of the cat states is negatively related with $\alpha$ as $\exp(-L\alpha ^2)$, where $L$ is determined by cavity loss. We find that $\alpha$-coherence is numerically sensitive to the off-diagonal elements of the density matrix, and thus the noticeable deviation between the theoretical curve and experimental data is attributed to the daily fluctuation in qubit coherence.

\subsection{Squeezing.}
Squeezing refers to a reduced quantum fluctuation for a certain quadrature of the photon field than that in vacuum state or coherent state, which is important for various applications in quantum metrology~\cite{Pooser19,Xiao19}.
Quantum interference of the coherent state components could result in squeezing of the reflected photon fields. 
To quantify the squeezing effect in the experimentally prepared states, we use a generalized squeezing parameter defined as~\cite{Knight1992} 
\begin{equation}
S_p^{(N)}=\frac{\langle (\Delta \hat{p})^N \rangle - C^{N/2}(N-1)!!}{C^{N/2}(N-1)!!},
\label{2ndS}
\end{equation}
where N is the order of squeezing, $C=1/4$ is the constant of commutation relation and $\langle (\Delta \hat{p})^N \rangle$ is the $N$th order moment of the quadrature operator in $p$-direction, which is the expected squeezing direction for the experimental configuration.
$S_p^{(N)}<0$ indicates the existence of squeezing, and the maximum squeezing corresponds to $S_p^{(N)}=-1$. We shall only consider the moments of even order since otherwise we would always have $S_p^{(N)}>0$~\cite{Knight1992}. 
Second-order squeezing has been extensively studied with a broad spectrum of quantum state, describing a smaller variance of the field distribution than that of a coherent state. 
Higher order squeezing with $N>2$ essentially describes a reduced quantum fluctuation for higher order moments of the photon field, allowing much fractional noise reduction than lower-order squeezing~\cite{Mandel1985}.

Previous pioneering works demonstrate that second-order squeezing can be observed in even cat state with small $\alpha$~\cite{Kien1991}. 
Fig.~\ref{quantum_coherence}(a) shows $S_p^{(2)}$ with varied $\alpha$ for the experimental states conditioned on the qubit state in $\ket{0}$, which corresponds to even cat states with varied sizes. Second-order squeezing is observed for all of the experimental states as expected. Moreover, a gradually declined squeezing level is observed in Fig.~\ref{quantum_coherence}(b) when the experimental state evolves from an even cat state to a coherent state, which is consistent with the theory. Additionally, when varying $\theta$ from 0 to $\pi$ while keeping $\xi=\pi/2$ (Fig.~\ref{quantum_coherence}(c)), we observe a continuous evolution from a squeezed even cat state to an anti-squeezed odd cat state. For the $S_p^{(2)}$ evolution mentioned above, theoretical results based on the error model are in agreement with the experimental results, with the deviations mainly originating from daily fluctuation in qubit coherence.

Prominently, we observe explicit fourth-order squeezing in the experimental states. As shown in Fig.~\ref{quantum_coherence}(b), $S_p^{(4)}$ shows negative values for the even-cat-like states with varied relative ratio between $\ket{\alpha}$ and $\ket{-\alpha}$. Also, in Fig.~\ref{quantum_coherence}(a), for even cat states with equal proportion of two coherent contributions, fourth-order squeezing can be observed with smaller $\alpha$. Those results demonstrate a versatile control on the squeezing of photon field with our method.

\section{Conclusion}

In conclusion, we have experimentally prepared various kinds of itinerant cat states in the microwave domain by reflecting coherent state photons from a cavity containing a superconducting qubit. 
The superposition of coherent components can be well controlled through a coherent control of the qubit state.
The preparation of generalized cat states is confirmed via a quantum state tomography on the reflected photon field, and further quantified in the frame of quantum coherence resource theory. We note that this method is of great versatility in controlling the statistical properties of a photon field, for which we have realized a continuous evolution of the reflected field from a super-Poissonian distribution to a sub-Poissonian distribution, or from anti-squeezing to squeezing. Even more, we have observed four-order squeezing in the prepared states. Our results demonstrate a powerful toolbox for many quantum information processing protocols based on coherent state superposition or non-classical light field statistics~\cite{vanLoock2008,Jiang2017prl,Munro1999,Kim2002,Ralph2003,Ralph2008,Gilchrist_2004,Sciarrino2020,Jiang2017prl,Jiang2017,Pooser19,Xiao19}.

\section{ACKNOWLEDGMENTS}

This work was supported by the Frontier Science Center for Quantum Information of the Ministry of Education of China through the Tsinghua University Initiative Scientific Research Program, the National Natural Science Foundation of China under Grant No.11874235 and No.11925404, the National key Research and Development Program of China (2017YFA0304303), the Key-Area Research and Development Program of Guangdong Province (No.2020B0303030001) and a grant (No.2019GQG1024) from the Institute for Guo Qiang, Tsinghua University. Y.K.W. acknowledges support from the start-up fund from Tsinghua University.

\appendix

\section{\label{sec:app-A} Sample and measurement setup}

The sample consists of a three-dimensional microwave cavity made from bulk aluminum, and a superconducting qubit made from aluminum film on a sapphire substrate~\cite{wang2021flying}. The qubit is placed at the center of the 3D cavity to achieve strong coupling. The out-coupling rate of the cavity is precisely tuned by adjusting the length of a one-dimensional transmission line extended into the cavity, to meet the optimal phase condition as $\kappa_{tot} \sim 2|\chi|$, as shown in Fig.~1(b) of the main text. The sample is cooled to about 20 mK in a dilution refrigerator for experiments. A detailed list of the device parameter can be found in Table \ref{table-1}.

\begin{figure*}[!tbp]
\centering
\includegraphics[width=1.1\linewidth]{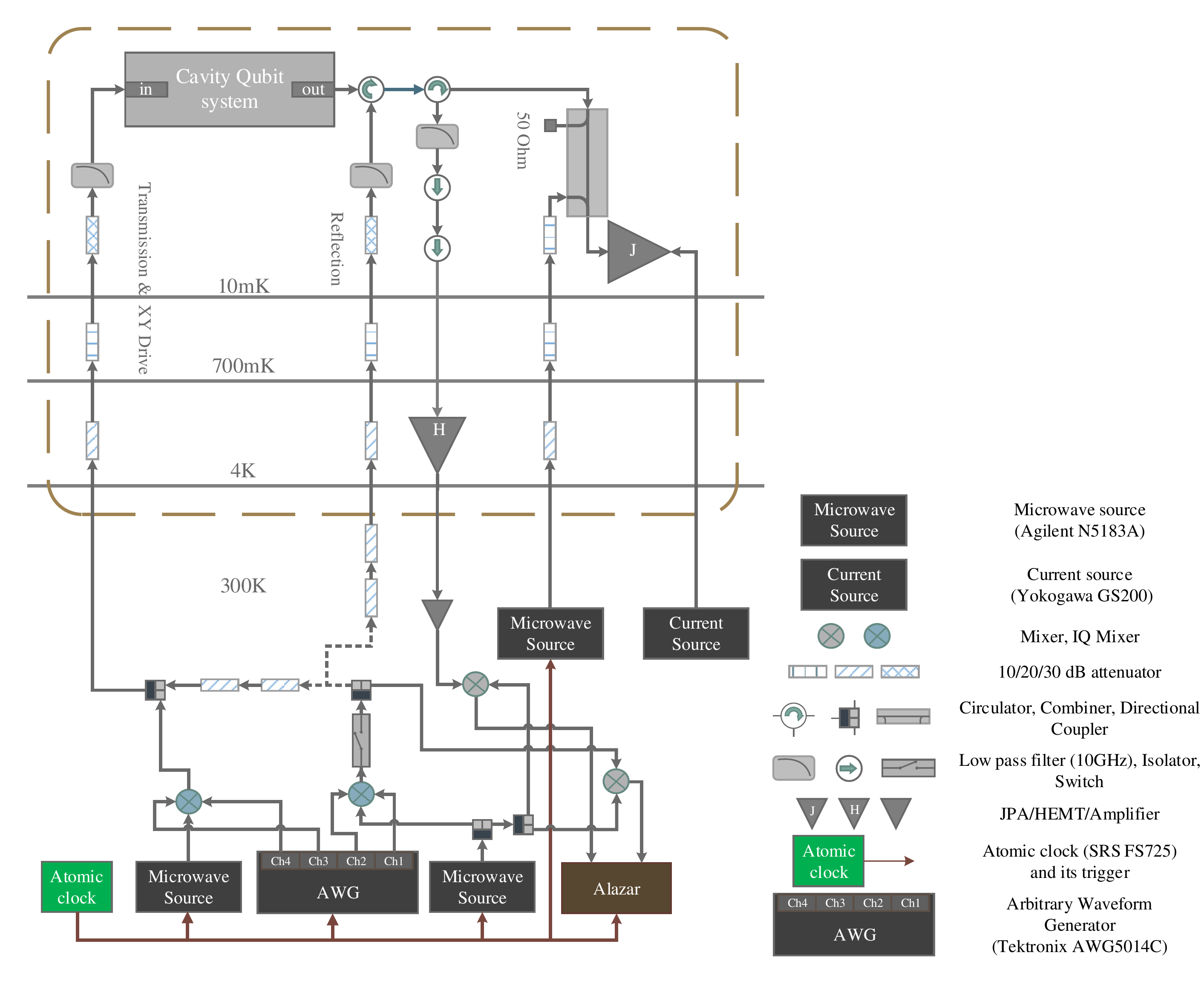}
\caption{\textbf{An illustration of the experimental setup.}}
\label{setup}
\end{figure*}

The measurement setup is schematically shown in Fig.~\ref{setup}~\cite{wang2021flying}. Specifically, the cavity reflection signal is successively amplified by a cascade amplifier circuit containing a Josephson parametric amplifier (JPA) at the base plate, a high-electron-mobility transistor (HEMT) amplifier at 4K plate and two microwave amplifiers at room temperature. The amplified signal is finally acquired by a homodyne setup. The JPA is working in a phase preserving mode with the gain of 16 dB around the cavity frequency, yielding an overall detection efficiency 1/($n_{noise}$+1) = 20\%, with a noise photon number $n_{noise}$=4. In this way, we measure the amplified in-phase and quadrature signals of the reflected photon field, which can be used for the calculation of moments for the photon distribution and quantum state tomography, as seen in the next section.

\section{\label{sec:app-B} Moments and quantum state tomography}

\subsection{\label{sec:app-B-1} Photon number calibration.}
In this cat state generation scheme, the size of the cat state is determined by the input coherent state of which the strength $\left|\alpha\right|^2$ is proportional to the input photon flux $\dot{n}_d$. In order to confirm the size of the cat state, photon flux $\dot{n}_d$ is calibrated with the method introduced in Ref.~\cite{Nakamura2018}. This method is based on the additional qubit dephasing rate $\Gamma_m$ induced by AC stark shift when the cavity is occupied by a certain number of photons. With a continuous coherent drive at frequency $\omega_d$ applied to the qubit-cavity system, the additional qubit dephasing rate $\Gamma_m$ can be expressed as
\begin{equation}
\begin{split}
\Gamma_m&=\frac{\kappa_{tot}\chi^2}{\kappa_{tot}^2/4+\chi^2+\Delta_d^2}(\bar{n}_++\bar{n}_-)\\
\bar{n}_\pm&=\frac{\kappa_r\dot{n_d}}{\kappa_{tot}^2/4+(\Delta_d\pm\chi)^2},
\label{drive-dephase}
\end{split}
\end{equation}
where $\bar{n}_\pm$ is the average photon number in the cavity when the qubit is in $\ket{0}$ or $\ket{1}$; The detuning between the coherent driving and cavity frequency $\Delta_d=\omega_d-\omega_c=-0.1\,$MHz is chosen during calibration. By fitting the relation between the practical qubit dephasing rate and varied cavity input signal strength with Eq.~\ref{drive-dephase}, the corresponding photon flux in the cat state preparation experiment is figured out as $\dot{n}_d=1.35\pm0.05\mu s^{-1}$. Then by using input-output theory~\cite{walls2008quantum}, the size of the prepared cat state in the reflection path $\left|\alpha\right|$ can be calculated as
\begin{equation}
|\alpha|=\sqrt{n_r}=|\frac{i\kappa_r}{\chi+i\kappa_{tot}/2}-1|\sqrt{\dot{n}_d T} .
\end{equation}
Here the drive signal is taken resonant with the cavity mode $\omega_d=\omega_c$, and $T$ is the length of the coherent pulse. In our experiment, with the length of the coherent pulse $T=1~\mu$s, the size of the cat state with varied $\xi$ and $\theta_q$ is fixed at $\left|\alpha\right|=1.07\pm0.04$, while the even/odd cat state with varied size $\left|\alpha\right|$ is investigated from 0.8 to 1.3.

\begin{table}
\caption{System parameters}
\begin{tabular}{cccccc}
\hline
\hline
cavity bare frequency $\omega_c/2\pi$(GHz) & 8.6885 \\
\hline
cavity internal loss rate $\kappa_i/2\pi$(MHz)& $0.22$ \\
\hline
cavity out-coupling rate $\kappa_r/2\pi$(MHz)  & $2.23$\\
\hline
qubit frequency $\omega_q/2\pi$(GHz) & 5.2927\\
\hline
dispersive coupling rate $\chi/2\pi$(MHz) & -1.1\\
\hline
qubit energy relaxation time $T_1(\mu s)$  & 20  \\ 
\hline
qubit dephasing time $T_2(\mu s)$ &6\\
\hline
qubit readout fidelity &$97.0\%$\\
\hline
JPA gain(dB)& 16\\
\hline
\hline
\end{tabular}
\label{table-1}
\end{table}

\subsection{\label{sec:app-B-2} Moments of the photon field and quantum state tomography.}

As discussed in last section, our measurement setup effectively performs a homodyne detection on the propagating photon modes~\cite{Eichler2011,Eichler2012}. The two conjugate quadratures $I,Q$ constitute the complex amplitude $S=I+iQ$, which is repeatedly recorded during the experiment. In principle the quantum state of the propagating light field can be fully reconstructed based on the distribution of the measured quadratures. A practical issue is that the measured quadratures contain both the signal of reflected photons and the noise added by the detection chain. Therefore one has to firstly extract the information of the reflected photon field from the measured quadratures for further state tomography. A detailed discussion about this method can be found in Ref.~\cite{Eichler2012}.

Considering both the amplified cavity reflection and the added noise, the measured complex amplitude $S$ is equivalent to the result of the measurement operator $\hat{S}=\hat{a}+\hat{h^\dagger}$, where $\hat{a}$ is the annihilation operator of the reflected microwave photon mode and $\hat{h^\dagger}$ is the creation operator of the noise mode, which originates from the added noises of the amplifiers and the circuit losses~\cite{Lehnert2011}. In order to separate the noise from the signal, the complex amplitude was measured for both the desired state and the vacuum state, with the measurement operator expressed as $\hat{S}_{sig}=a+h^\dagger$ and $\hat{S}_{vac}=h^\dagger$ respectively. Assuming the noise added by the detection chain is independent of the photon signal, there is no correlation between the signal mode $a$ and noise mode $h$. This means the moments $\langle (\hat{S}^\dagger_{sig})^m\hat{S}^n_{sig}\rangle$ of the complex amplitude with signal, $S_{sig}$, can be expanded as
\begin{equation}
\langle(\hat{S}^\dagger_{sig})^m\hat{S}^n_{sig}\rangle=\sum_{i,j=0}^{m,n}\binom{n}{j}\binom{m}{i}\langle(\hat{a}^\dagger)^i\hat{a}^j\rangle\langle\hat{h}^{m-i}(\hat{h}^\dagger)^{n-j}\rangle,
\end{equation}
where $\langle\hat{h}^{m-i}(\hat{h}^\dagger)^{n-j}\rangle$ corresponds to the moment of measured vacuum state $\langle(S_{vac}^\dagger)^{m-i}S_{vac}^{n-j}\rangle$. By solving these equations, the moments of only the signal photon state without noise $\langle(\hat{a}^\dagger)^m\hat{a}^n\rangle$ can be obtained.

Further, based on the measured moments, one could reconstruct the density matrix of the itinerant photon field with quantum state tomography~\cite{Eichler2012}. With the moments in different orders $\langle(\hat{a}^\dagger)^m\hat{a}^n\rangle$ of the measured photon state and its standard deviations $\delta_{m,n}$, a most likely density matrix of the photon state can be obtained by applying a maximum likelihood method. The log-likelihood function is shown as:
\begin{equation}
L_{log}=-\sum_{n,m}\frac{1}{\delta_{m,n}^2}|\langle(\hat{a}^\dagger)^m\hat{a}^n\rangle-\Tr[\rho_{ph}(\hat{a}^\dagger)^m\hat{a}^n]|^2,
\end{equation}
Maximizing this log-likelihood function with the physical constraints $\rho_{ph}\ge 0$ and $\Tr\rho_{ph}=1$, the experiment state can be consequently reconstructed. In our experiment, the complex amplitudes of the amplified propagating mode are sampled by $3\times10^7$ times. The moments up to 6 order ($m+n\leq6$) were taken into consideration, and a cut-off photon number of 11 is used for the reconstruction of the quantum states.

\section{\label{sec:app-C} $\alpha$-coherence.}
The generalized cat states realized in the experiment consist of various kinds of superpositions of two coherent states. In order to quantify the quantum coherence in the cat states, especially in the presence of loss, we calculate the $\alpha$-coherence of the prepared states.

As described in Ref.~\cite{Plenio2014}, the original proposal of the coherence in a certain quantum state is discussed in the discrete finite dimensional case, which is far cry from the cat states studied here. Taking the Fock basis $\{\ket{n}\}$ to illustrate, the two coherent state components would have non-zero coherence even larger than that of their superpositions, which indicates that this quantifier is incongruent to quantify the superposition induced quantum coherence here.
A recent work has developed an approach to extend the application of the existing quantifiers to the arbitrary superposition of coherent states, through a proper orthonormalization on the basis set and state decomposition~\cite{Jeong2017}.
First, the considered system $A$ is jointed with an $N+1$ dimensional auxiliary system $B$ with an orthonormal basis set $\{\ket{i}_{B}\}$ ($i=0,\cdots,N$), expanding the target density matrix $\rho_{A}$ to a tensor product $\rho^{(0)}_{AB}=\rho_{A}\otimes\ket{0}_B\bra{0}$. Then, one could figure out a set of coherent states $\ket{\alpha^{(i)}}_{A}$, achieving the condition
 \begin{equation}
\begin{aligned}
&\Tr(\ket{\alpha^{(i)}}_{A}\bra{\alpha^{(i)}}\otimes\ket{0}_{B}\bra{0}\rho^{(i-1)}_{AB}) \\
&=\max_\alpha{\Tr(\ket{\alpha}_{A}\bra{\alpha}\otimes\ket{0}_{B}\bra{0}\rho^{(i-1)}_{AB})},
\label{alpha condition}
\end{aligned}
\end{equation} 
where $\rho^{(i)}_{AB}\equiv U_{\alpha^{(i)}}\rho_{AB}^{(i-1)}U_{\alpha^{(i)}}^{\dagger}$ and $U_{\alpha^{(i)}}\equiv I\otimes I+\ket{\alpha^{(i)}}_{A}\bra{\alpha^{(i)}}\otimes(\ket{i}_{B}\bra{0}+\ket{0}_{B}\bra{i}-\ket{0}_{B}\bra{0}-\ket{i}_{B}\bra{i})$. After $N$ unitary transformations, the population originally concentrated in the subspace $\ket{0}_{B}\bra{0}$ is spread around the total space with orthogonal components $\{\ket{\alpha^{(i)}}_{A}\ket{i}_{B}\}$, leaving an almost zero trace in the subspace $\ket{0}_{B}\bra{0}$. By projecting the transformed density matrix $\rho^{(N)}_{AB}$ onto the subspace $\{\ket{\alpha^{(i)}}_{A}\ket{i}_{B}\}$ with projector $\Pi^{(N)}\equiv\sum^{N}_{i=1}\ket{\alpha^{(i)}}_{A}\bra{\alpha^{(i)}}\otimes\ket{i}_{B}\bra{i}$, the original density matrix is reconstructed on a coherent state basis set as $\rho_{\alpha}=\mathscr{N}(\Pi^{(N)}\rho^{(N)}_{AB}\Pi^{(N)})$, which is now compatible with the commonly used coherence quantifiers~\cite{Plenio2014}. The amount of coherence calculated in this density matrix form is called $\alpha$-coherence. Here we employ the relative entropy as quantifier to meter quantum coherence, defining a $C_{\alpha}$ as 
\begin{equation}
C_{\alpha}(\hat{\rho_\alpha})\equiv S(\hat{\rho}_{\alpha}^{diag})-S(\hat{\rho}),
\label{relative entropy}
\end{equation}
where $S(\hat{\rho})=-\Tr(\hat{\rho}\log_2\hat{\rho})$ is the von Neumann entropy.

\section{\label{sec:app-D} The loss model and error budget} 

\subsection{\label{sec:app-D-1} The loss model }

In this part, the contribution of cavity loss, finite qubit lifetime ($T_1,T_2$) and qubit state measurement error to the infidelity of the generalized cat states is calculated.


Considering finite cavity loss, the qubit-photon entangled system can be written as $(\ket{0}\ket{\alpha_0^r}\ket{\alpha_0^l}+\ket{1}\ket{\alpha_1^r}\ket{\alpha_1^l})/\sqrt{2}$. Using input-output theory~\cite{Gardiner1985,walls2008quantum}, the reflection mode $\ket{\alpha_{0/1}^r}$ and the loss mode $\ket{\alpha_{0/1}^l}$ conditioned on qubit state $\ket{0}(\ket{1})$ can be written as 

\begin{equation}
\begin{split}
\alpha_{0/1}^l&=\frac{i\sqrt{\kappa_r\kappa_i}}{\Delta\pm\chi+i\kappa_{tot}/2}\alpha_{in},\\
\alpha_{0/1}^r&=(\frac{i\kappa_r}{\Delta\pm\chi+i\kappa_{tot}/2}-1)\alpha_{in}
\end{split}
\end{equation}

where $\alpha_{in}$ is the amplitude of input coherent state photon, $\Delta$ is detuning between the frequency of the input photon and the cavity bare frequency, $\kappa_{r}$ is the out-coupling rate of cavity at the reflection port, and $\kappa_{tot}=\kappa_i+\kappa_r$ is the total linewidth of the cavity mode with $\kappa_i$ the internal loss rate of the cavity. In the experiment, in order to engender a $\pi$ phase shift between $\alpha_0^r$ and $\alpha_1^r$ when cavity is driven resonant $\Delta=0$, $\kappa_r$ was fine-tuned to meet the condition, $\kappa_r^2-\kappa_i^2=4\chi^2$. This condition can also be rewritten as $\kappa_{tot}\approx2\chi$, because $\kappa_i\ll\kappa_r$.  According to our experimental setup, the amplitude of the reflection mode and the loss mode is expressed as

\begin{equation}
\begin{split}
\alpha_{0/1}^r&=\pm\eta\alpha_{in}\equiv\alpha\\
\alpha_{0/1}^l&=(1\pm i\eta)\sqrt{\frac{1-\eta^2}{1+\eta^2}}\alpha_{in}\equiv (1/\eta\pm i)\sqrt{\frac{1-\eta^2}{1+\eta^2}}\alpha
\end{split}
\end{equation}
where $\eta=\sqrt{(1-\kappa_i/\kappa_r)/(1+\kappa_i/\kappa_r)}$. The photon in the loss mode shall be traced off from the system. After applying an arbitrary rotation $R_{\theta_q}(\xi)=\exp(-i\frac{\xi}{2}(\sigma_x\sin\theta_q-\sigma_y\cos\theta_q))$, the measured photon state for qubit at $\ket{0}$ and $\ket{1}$ would be

\begin{equation}
\begin{split}
\rho_0^0&=\mathscr{N}(\cos^2\frac{\xi}{2}\ket{\alpha}\bra{\alpha}+\cos\frac{\xi}{2}\sin\frac{\xi}{2}e^{-i\theta_q}\bra{\alpha_0^l}\ket{\alpha_1^l}\ket{-\alpha}\bra{\alpha}\\
&+\cos\frac{\xi}{2}\sin\frac{\xi}{2}e^{i \theta_q}\bra{\alpha_1^l}\ket{\alpha_0^l}\ket{\alpha}\bra{-\alpha}+\sin^2\frac{\xi}{2}\ket{-\alpha}\bra{-\alpha}),\\
\rho_1^0&=\mathscr{N}(\sin^2\frac{\xi}{2}\ket{\alpha}\bra{\alpha}-\cos\frac{\xi}{2}\sin\frac{\xi}{2}e^{-i\theta_q}\bra{\alpha_0^l}\ket{\alpha_1^l}\ket{-\alpha}\bra{\alpha}\\
&-\cos\frac{\xi}{2}\sin\frac{\xi}{2}e^{i \theta_q}\bra{\alpha_1^l}\ket{\alpha_0^l}\ket{\alpha}\bra{-\alpha}+\cos^2\frac{\xi}{2}\ket{-\alpha}\bra{-\alpha}),
\label{loss_state}
\end{split}
\end{equation}
Comparing with ideal cat state in Eq.~2 in the main text, a decoherence factor $\bra{\alpha_{1/0}^l}\ket{\alpha_{0/1}^l}$ appears as
\begin{equation}
\bra{\alpha_{0/1}^l}\ket{\alpha_{1/0}^l}=\exp(-2\frac{1-\eta^2}{1+\eta^2}(\eta^2\pm i\eta)\frac{|\alpha|^2}{\eta^2}).
\label{dc-fac}
\end{equation}
The finite internal loss of cavity gives rise to a certain decoherence in the superposition of the two coherent state $\ket{\pm\alpha}$. This undesired decoherence has a negative exponential relation with the cat state size $\alpha$, imposing a severe restrictions on the preparation fidelity of large cat state. 
Additionally, by Eq.~\ref{dc-fac} and Eq.~\ref{loss_state}, there would be an azimuthal angle deviation $\delta\theta=2\frac{1-\eta^2}{1+\eta^2}\frac{|\alpha|^2}{\eta}$ induced by cavity loss, which is also confirmed by the experiment. This angle deviation can be compensated by setting an offset for $\theta$ in the arbitrary rotation $R_{\theta}(\xi)$ to $\theta - \theta_q=\delta\theta = 0.125\pi$, where $\theta$ essentially refers to the superposition phase of the coherent states.

The effect of finite lifetime of qubit on the qubit-photon entanglement state can be counted by using master equation with Lindblad operators: $L_{T_1}=\frac{1}{\sqrt{T_1}}\ket{0}\bra{1}$ for qubit decay, and $L_{T_\phi}=\frac{1}{\sqrt{2T_\phi}}(\ket{0}\bra{0}-\ket{1}\bra{1})$ for qubit dephasing. Here  $T_{\phi}=(\frac{1}{T_2}-\frac{1}{2T_1})^{-1}$ is the pure dephasing time in total $T_2$ contribution. By integrating the master equation over the duration time of the full experimental sequence $t$, the density matrix of the qubit-photon entanglement system can be written as  



\begin{equation}
\begin{split}
\rho_{q-ph}^{LT}=&\frac{1}{2}\ket{0}\bra{0}\otimes(\left(1-\exp(-\frac{t}{T_1})\right)\ket{-\alpha}\bra{-\alpha}+\ket{\alpha}\bra{\alpha})\\
+&\frac{1}{2}\bra{\alpha_{1}^l}\ket{\alpha_{0}^l}\exp(-\frac{t}{T_2})\ket{0}\bra{1}\otimes\ket{\alpha}\bra{-\alpha} \\
+&\frac{1}{2}\bra{\alpha_{0}^l}\ket{\alpha_{1}^l}\exp(-\frac{t}{T_2})\ket{1}\bra{0}\otimes\ket{-\alpha}\bra{\alpha}\\
+&\frac{1}{2}\exp(-\frac{t}{T_1})\ket{1}\bra{1}\otimes\ket{-\alpha}\bra{-\alpha}
\end{split}
\label{q-ph-LT}
\end{equation}

After applying an arbitrary rotation $R_{\theta}(\xi)$, the measured photon state for qubit at $\ket{0}$ and $\ket{1}$ would be

\begin{equation}
\begin{split}
\rho_0^{LT}&=\mathscr{N}(\cos^2\frac{\xi}{2}(\left(1-\exp(-\frac{t}{T_1})\right)\ket{-\alpha}\bra{-\alpha}+\ket{\alpha}\bra{\alpha})\\
&+\cos\frac{\xi}{2}\sin\frac{\xi}{2}e^{-i\theta}\exp(-\frac{t}{T_2})\bra{\alpha_0^l}\ket{\alpha_1^l}\ket{-\alpha}\bra{\alpha}\\
&+\cos\frac{\xi}{2}\sin\frac{\xi}{2}e^{i \theta}\exp(-\frac{t}{T_2})\bra{\alpha_1^l}\ket{\alpha_0^l}\ket{\alpha}\bra{-\alpha}\\
&+\sin^2\frac{\xi}{2}\exp(-\frac{t}{T_1})\ket{-\alpha}\bra{-\alpha}),\\
\rho_1^{LT}&=\mathscr{N}(\sin^2\frac{\xi}{2}(\left(1-\exp(-\frac{t}{T_1})\right)\ket{-\alpha}\bra{-\alpha}+\ket{\alpha}\bra{\alpha})\\
&-\cos\frac{\xi}{2}\sin\frac{\xi}{2}e^{-i\theta}\exp(-\frac{t}{T_2})\bra{\alpha_0^l}\ket{\alpha_1^l}\ket{-\alpha}\bra{\alpha}\\
&-\cos\frac{\xi}{2}\sin\frac{\xi}{2}e^{i \theta}\exp(-\frac{t}{T_2})\bra{\alpha_1^l}\ket{\alpha_0^l}\ket{\alpha}\bra{-\alpha}\\
&+\cos^2\frac{\xi}{2}\exp(-\frac{t}{T_1})\ket{-\alpha}\bra{-\alpha}),
\label{loss_decay_state}
\end{split}
\end{equation}

\begin{figure*}[hbtp]
\centering
\includegraphics[width=0.8\linewidth]{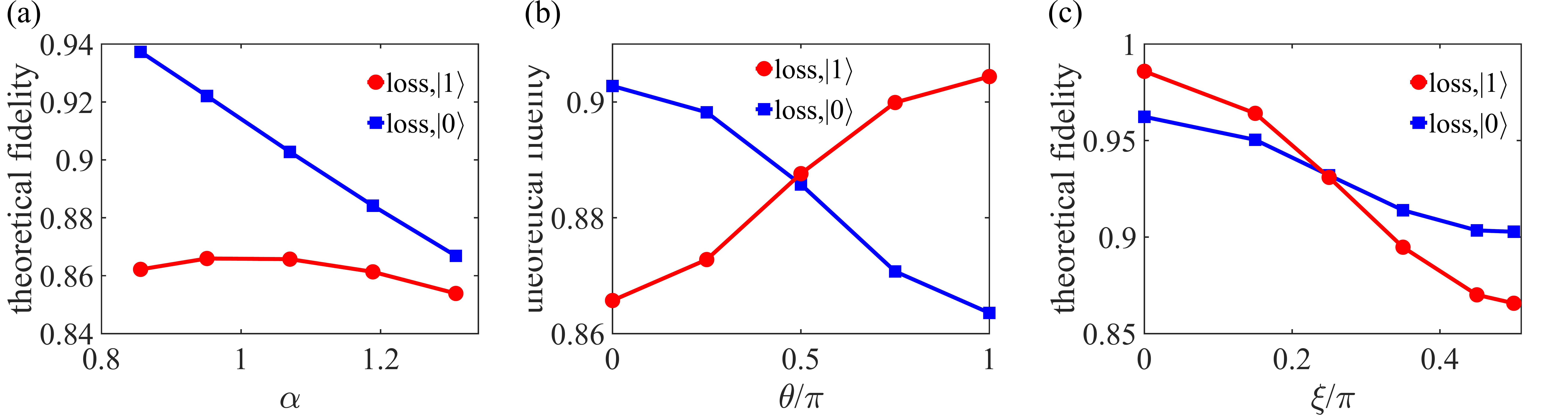}
\caption{\textbf{Fidelities of theoretical prediction for the generalized cat states.} The theoretically predicted fidelities for the generalized cat states conditioned on the qubit state either in $\ket{0}$ or in $\ket{1}$, with (a) different coherent state amplitude $\alpha$ and fixed $\xi=\pi/2$ and $\theta=0$, (b) different superpostion phase $\theta$ when using $\alpha=1.07$ and $\xi=\pi/2$ and (c) different population fraction of the two coherent components $\xi$ while keeping $\alpha=1.07$ and $\theta=0$. The red circles and the blue squares on lines are the theoretical results corresponding to the particularly prepared states in the experiment.}
\label{theoretical fidelity}
\end{figure*}

Since each generalized cat state is prepared with projecting the final qubit-photon entanglement system on a certain qubit state $\ket{0}$ or $\ket{1}$, wrong attribution of the qubit state will mix the cat states conditioned on $\ket{0}$ and $\ket{1}$, and add an infidelity of measurement error. For a qubit with probability $P_{0/1}$ at state $\ket{0}(\ket{1})$, the measurement error $\epsilon_{0/1}$ leads to a wrong count probability $P_{0/1}\epsilon_{0/1}$ leaving the right count probability $P_{0/1}(1-\epsilon_{0/1})$. Using Bayes theory, the measured states conditioned on both $\ket{0}$ and $\ket{1}$ become mixtures as  
\begin{equation}
\begin{split}
\rho_0&=\frac{P_0(1-\epsilon_0)\rho_0^{LT}+P_1\epsilon_1\rho_1^{LT}}{P_0(1-\epsilon_0)+P_1\epsilon_1},\\
\rho_1&=\frac{P_1(1-\epsilon_1)\rho_1^{LT}+P_0\epsilon_0\rho_0^{LT}}{P_1(1-\epsilon_1)+P_0\epsilon_0}.
\end{split}
\label{lo_de_me}
\end{equation}
where $P_{0/1}$ is the probability for the qubit-photon entanglement system being projected to qubit state $\ket{0}(\ket{1})$, which can be obtained by the normalization factor in Eq.\ref{loss_decay_state}. With the three error factors considered, Eq.~\ref{lo_de_me} gives the theoretically predicted states, and is used to calculate the dashed lines in the figures for comparisons with experimental data.

\subsection{\label{sec:app-D-2} Error budget}

It is then possible to perform an explicit error budget for the experiments based on the error model developed before. 
In Fig.~\ref{theoretical fidelity}, we show the theoretical predicted fidelity for the generalized cat states. Owing to cavity loss and finite qubit lifetime, the fidelity shows a clear decreasing trend with $\alpha$ and $\xi$ increasing for the generalized cat states conditioned on qubit state in both $\ket{0}$ and $\ket{1}$. This feature is confirmed by the experimental data shown in the main text and in Fig.~\ref{vareidalpha}, which indicates that the loss model covers the main sources of errors in the experiment. It is interesting to find that in general, the odd cat states suffer more errors than the even cat states, which is detailed discussed in Ref.~\cite{wang2021flying}.

\begin{figure*}[!tbp]
\centering
\includegraphics[width=0.8\linewidth]{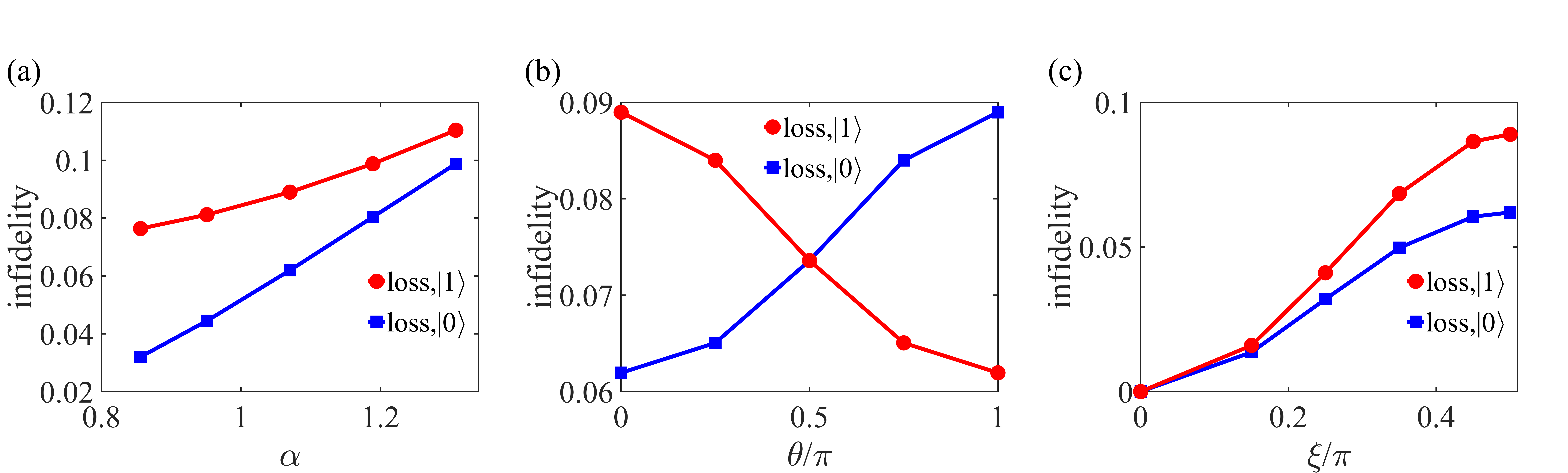}
\caption{\textbf{Cavity loss induced infidelities for the generalized cat states.} Cavity loss induced infidelities predicted according to Eq.~\ref{dc-fac} for the generalized cat states conditioned on the qubit state either in $\ket{0}$ or in $\ket{1}$, with (a) different coherent state amplitude $\alpha$ and fixed $\xi=\pi/2$ and $\theta=0$, (b) different superpostion phase $\theta$ when using $\alpha=1.07$ and $\xi=\pi/2$ and (c) different population fraction of the two coherent components $\xi$ while keeping $\alpha=1.07$ and $\theta=0$.}
\label{theoretical fidelity cavity}
\end{figure*}

\begin{figure*}[!tbp]
\centering
\includegraphics[width=0.8\linewidth]{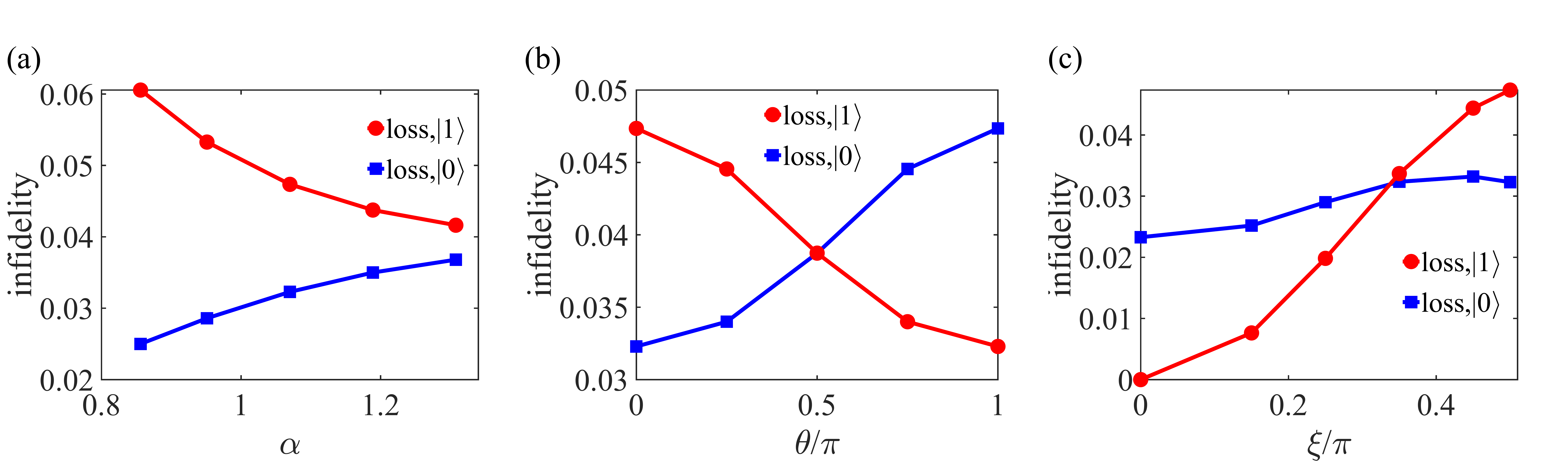}
\caption{\textbf{Qubit decay and dephasing induced infidelities for the generalized cat states.} Qubit decay and dephasing induced infidelities predicted according to Eq.~\ref{loss_decay_state} for the generalized cat states conditioned on the qubit state either in $\ket{0}$ or in $\ket{1}$, with (a) different coherent state amplitude $\alpha$ and fixed $\xi=\pi/2$ and $\theta=0$, (b) different superpostion phase $\theta$ when using $\alpha=1.07$ and $\xi=\pi/2$ and (c) different population fraction of the two coherent components $\xi$ while keeping $\alpha=1.07$ and $\theta=0$.}
\label{theoretical fidelity qubit}
\end{figure*}

\begin{figure*}[!tbp]
\centering
\includegraphics[width=0.8\linewidth]{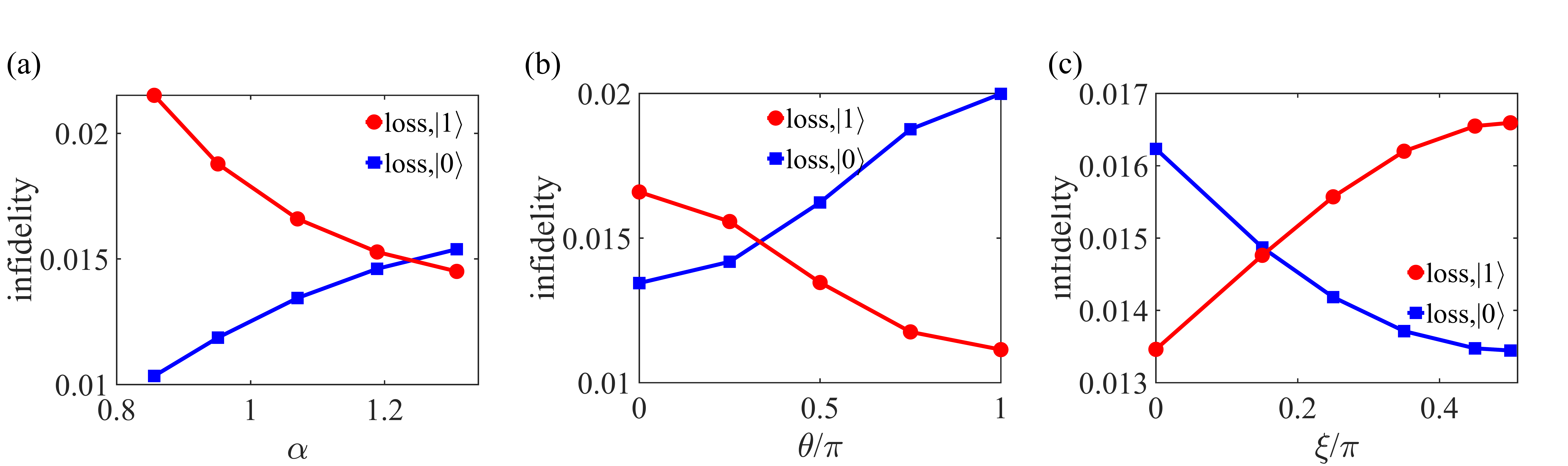}
\caption{\textbf{Qubit state readout error induced infidelities for the generalized cat states.} Qubit state readout error induced infidelities calculated with Eq.~\ref{lo_de_me} for the generalized cat states conditioned on the qubit state either in $\ket{0}$ or in $\ket{1}$, with (a) different coherent state amplitude $\alpha$ and fixed $\xi=\pi/2$ and $\theta=0$, (b) different superpostion phase $\theta$ when using $\alpha=1.07$ and $\xi=\pi/2$ and (c) different population fraction of the two coherent components $\xi$ while keeping $\alpha=1.07$ and $\theta=0$.}
\label{theoretical fidelity readout}
\end{figure*}

We could also separately consider the three error sources, including cavity loss, qubit decay/dephasing and qubit state readout error, to account for the preparation infidelity for the cat states, as shown in Fig.~\ref{theoretical fidelity cavity}, Fig.~\ref{theoretical fidelity qubit} and Fig.~\ref{theoretical fidelity readout}. The calculation is performed based on Eq.~\ref{dc-fac}, Eq.~\ref{loss_decay_state} and Eq.~\ref{lo_de_me} respectively, with an ideal state as initial state. It is worth mentioning that the calculation of total infidelity shown in Fig.~\ref{theoretical fidelity} combines all the three error sources together, and thus the total infidelity is smaller than the sum of the separated infidelities listed in Fig.~\ref{theoretical fidelity cavity}, Fig.~\ref{theoretical fidelity qubit} and Fig.~\ref{theoretical fidelity readout}, due to the fact that the infidelity induced by a specific error source for a partially mixed state is smaller than that for an ideal state.

Fig.~\ref{theoretical fidelity cavity} shows the cavity loss induced infidelity on the superposition states. One could find that with an increasing cat size $\alpha$, cavity loss induced infidelity rises quickly from a few percent to above ten percent, which is not surprising considering the fact that this infidelity scales with $\alpha$ as $\exp(-L\alpha ^2)$ in Eq.~\ref{dc-fac}. Therefore it is crucial to reduce cavity loss if a large cat size is desired. From Fig.~\ref{theoretical fidelity cavity}(c), one could find that the cavity loss related infidelity is positively related with the weight of the coherent state superposition, which means a superposition state with more quantum coherence (Fig.~3(b) in the main text) is more vulnerable to cavity loss.

Fig.~\ref{theoretical fidelity qubit} shows the qubit decay and dephasing induced infidelity on the coherent-state superposition. The qubit state induced infidelity is dominated by qubit dephasing, since T2 is closer to the experimental sequence duration and much shorter than T1 (see Table~\ref{table-1}). One could find that the qubit state related infidelity is also positively related with the weight of the coherent state superposition $\xi$, which is reasonable considering that qubit dephasing leads to the dephasing of coherent-state superposition in Eq.~\ref{loss_decay_state}. In our experiment, the qubit state caused about half the error as the cavity loss, which would get smaller with an increasing cat size $\alpha$.

Fig.~\ref{theoretical fidelity readout} shows the qubit state measurement error induced infidelity on the coherent-state superposition. Compared with the previous two error sources, qubit state measurement error induced infidelity varies within a relatively small region, which shows minor sensitivity to either $\alpha$ or $\xi$, which shall be greatly relieved by improving the qubit state readout performance.

\section{\label{sec:app-E} More data}

\subsection{\label{sec:app-E-1} Varied $\alpha$}

\begin{figure}[!tbp]
\centering
\includegraphics[width=1\linewidth]{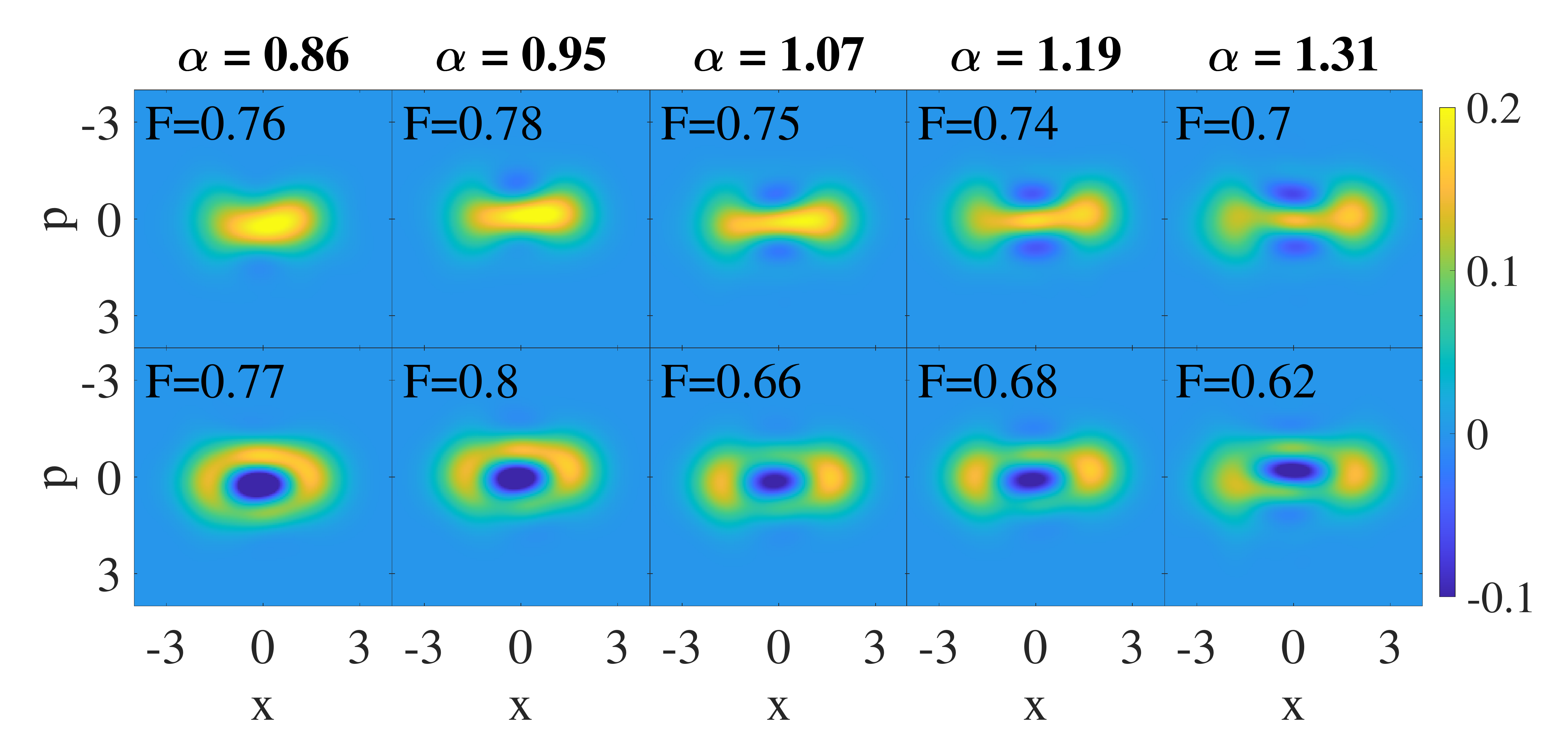}
\caption{\textbf{Winger function of the experimental cat states with varied sizes.} The reconstructed Wigner function for the reflected photon states conditioned on the qubit state in $\ket{0}$ and $\ket{1}$, where $\theta=0$ and $\xi=\pi/2$ are used. The amplitude of the coherent state $\alpha$ is varied from 0.86 to 1.31, as labeled on top of each column of the plots.}
\label{vareidalpha}
\end{figure}

The size of the cat state can be controlled by using different input coherent state $\alpha$. In Fig.~\ref{vareidalpha} we show the reconstructed Wigner function of odd and even cat states with varied $\alpha$ from 0.86 to 1.31. The fidelity of experimentally generated states shows a statistical decrease with an increasing $\alpha$, which is attributed to the cavity loss induced exponential decay term $\exp(-L\alpha ^2)$ in the density matrix (see Appendix ~\ref{sec:app-D-1} for details).

\subsection{\label{sec:app-E-2} Varied optical phase}

\begin{figure}[!tbp]
\centering
\includegraphics[width=0.9\linewidth]{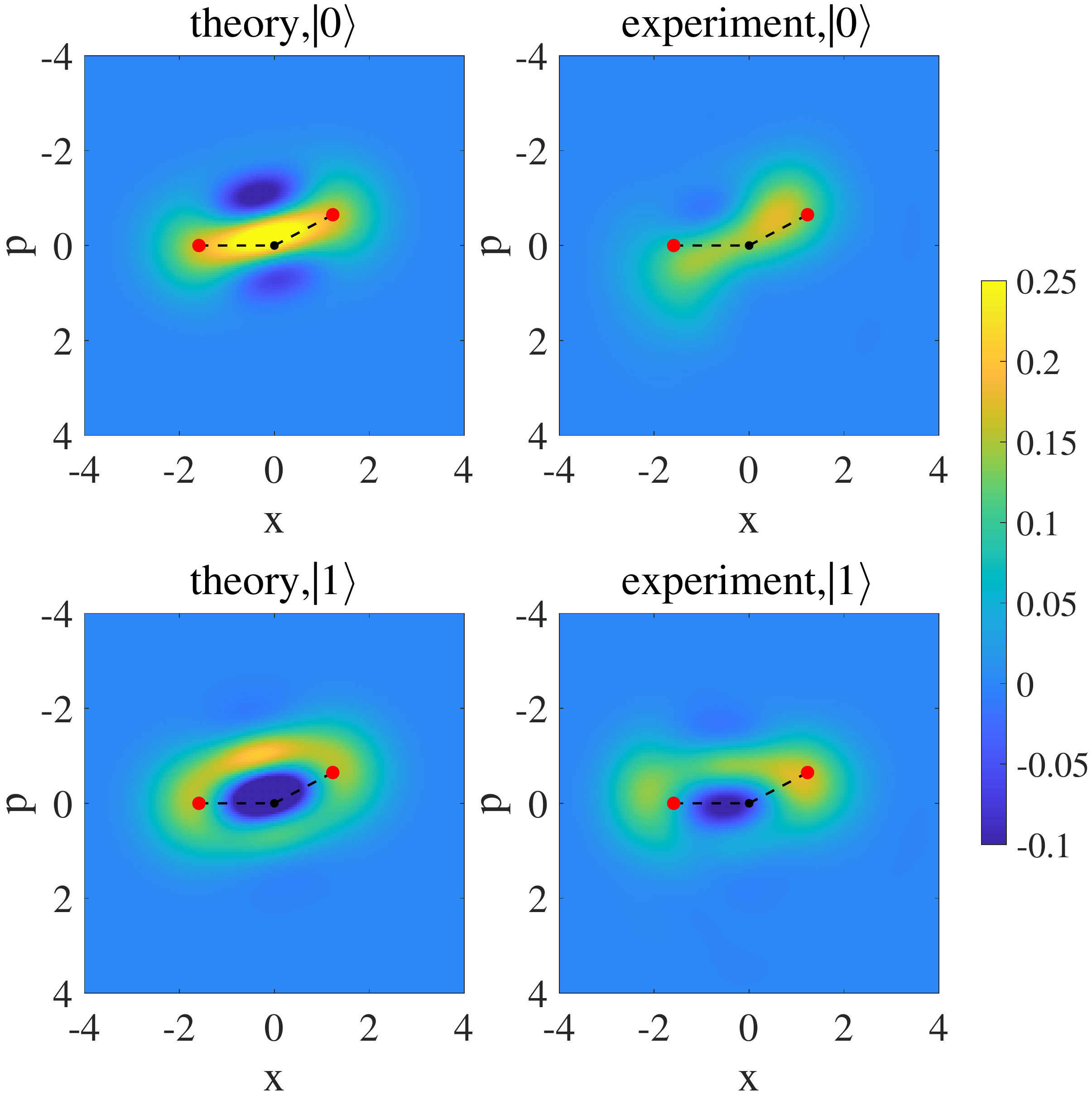}
\caption{\textbf{Winger function of the cat states generated with detuned microwave pulse from the cavity bare frequency.} The theoretical and experimentally reconstructed Wigner function for the reflected photon states conditioned on the qubit state in $\ket{0}$ and $\ket{1}$, where $\alpha=1.07$, $\theta=0$ and $\xi=\pi/2$ are used. The frequency of the input microwave photon pulse $\omega_p$ is detuned from the cavity bare frequency by $0.7$ MHz, with $\omega_p/2\pi-\omega_c/2\pi = 0.7$ MHz, corresponding to an optical phase difference of $2.657$ in radian between the two coherent state components. The black solid point in each panel is the original point. The red points indicate the centers of the two coherent components, from where a smaller optical phase difference than $\pi$ can be clearly seen.}
\label{vareidphase}
\end{figure}

The optical phase difference of coherent components in the superposition states is not limited to be $\pi$, but can be varied by using a different signal frequency. In the experiment, we tune the linewidth of the cavity to fit the dispersive shift as $\kappa_{tot}\approx2|\chi|$, which results in the phase difference between the reflectivities conditioned on the qubit state in $\ket{0}$ and in $\ket{1}$, as shown in Fig.~1(c) of the main text. For the generalized cat states shown in Fig.~2 of the main text and in Fig.~\ref{vareidalpha}, the signal frequency is tuned to the cavity bare frequency $\omega_p = \omega_c$, and thus we have two coherent components with the same amplitude but the opposite phases, as indicated by the dashed line in Fig.~1(b) and (c) of the main text. The optical phase difference can be controlled by using a detuned signal frequency from the cavity bare frequency $\omega_{c}$. As a proof of principle, we use a frequency detuning $\omega_p/2\pi-\omega_c/2\pi = 0.7$ MHz between the input signal frequency and cavity bare frequency in the experiment,  
which corresponds to an optical phase difference of $2.657$ in radian measured from Fig.~1(c) in the main text. The Wigner functions of the experimental states and the corresponding theoretical states are present in Fig.~\ref{vareidphase}. The centers of their coherent state components are labeled with red points in each panel, which indicate a well-tuned optical phase difference between the coherent states.

\subsection{\label{sec:app-E-3} Photon number statistic}
\begin{figure}[!tbp]
\centering
\includegraphics[width=0.9\linewidth]{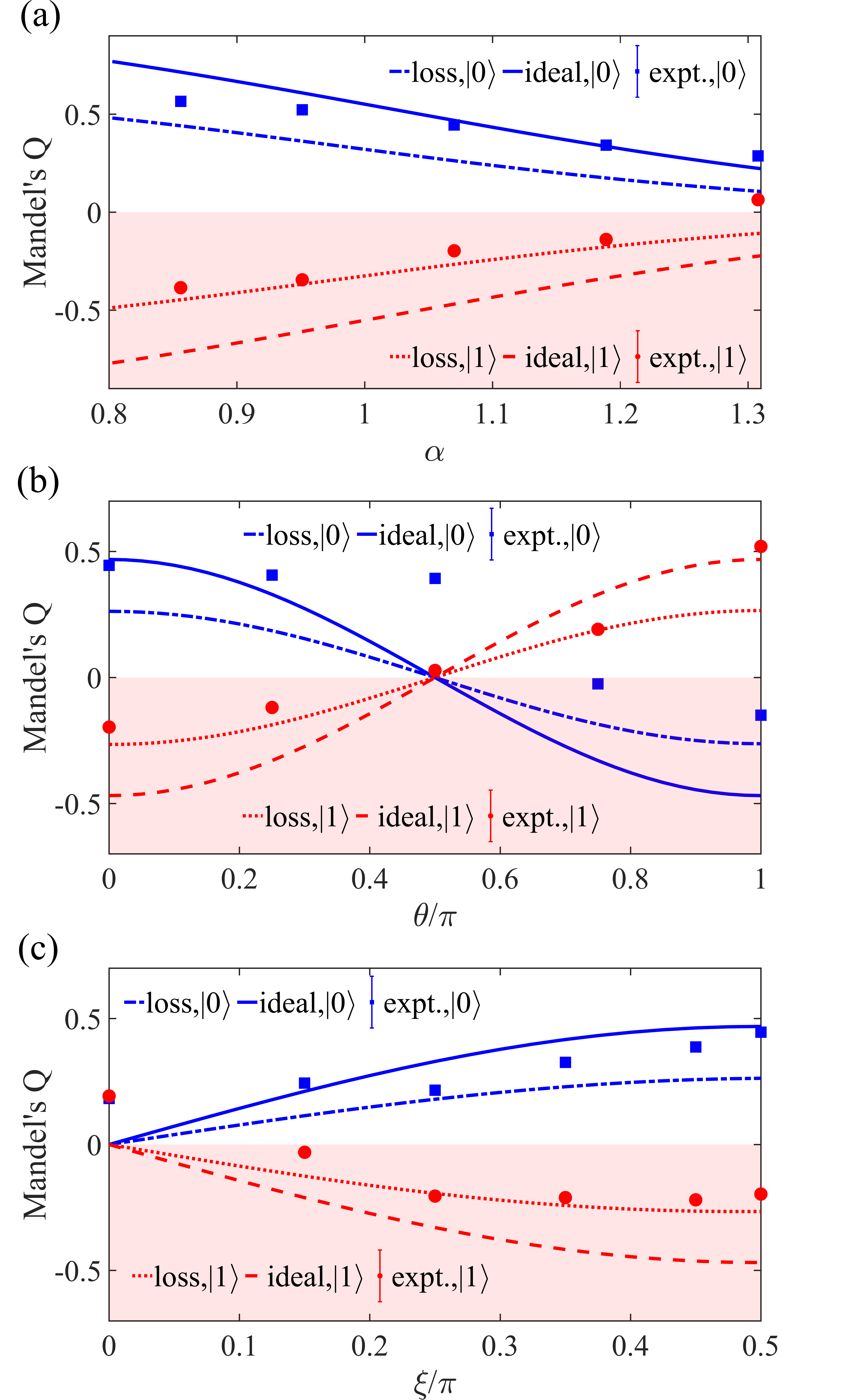}
\caption{\textbf{The Mandel's Q parameters for the prepared generalized cat states.} The calculated Mandel's Q parameters for the coherent state superpositions conditioned on the qubit state either in $\ket{0}$ or in $\ket{1}$, with (a) different coherent state amplitude $\alpha$ while using $\xi=\pi/2$ and $\theta=0$, (b) different superposition phase $\theta$ when using $\alpha=1.07$ and $\xi=\pi/2$ or (c) different population fraction of the two coherent components $\xi$ when using $\alpha=1.07$ and $\theta = 0$. The blue squares and the red circles show the experimental data conditioned on the qubit state in $\ket{0}$ and $\ket{1}$, respectively. The blue solid lines and the red dashed lines are theoretical results based on the corresponding ideal states conditioned on the qubit state in $\ket{0}$ and $\ket{1}$, respectively. The blue dotted dashed lines and the red dotted lines are theoretical results considering possible experimental loss and decoherence conditioned on the qubit state in $\ket{0}$ and $\ket{1}$, respectively. Q$<$0 indicates a sub-Poissonian photon distribution.}
\label{QQQ}
\end{figure}

Mandel's Q parameter is a direct measurement of the photon statistics, which is defined as~\cite{Mandel79}
 \begin{equation}
Q = \frac{\langle (\Delta \hat{n})^2 \rangle - \langle \hat{n} \rangle}{\langle \hat{n} \rangle}.
\label{Qpara}
\end{equation}
In the experiment, we calculate the Q parameters from the measured moments of the reflected photon states. In Fig.~\ref{QQQ} we systematically present the derived Mandel's Q parameter and photon number distribution for the experimentally prepared generalized cat states.

\begin{figure}[!tbp]
\centering
\includegraphics[width=1\linewidth]{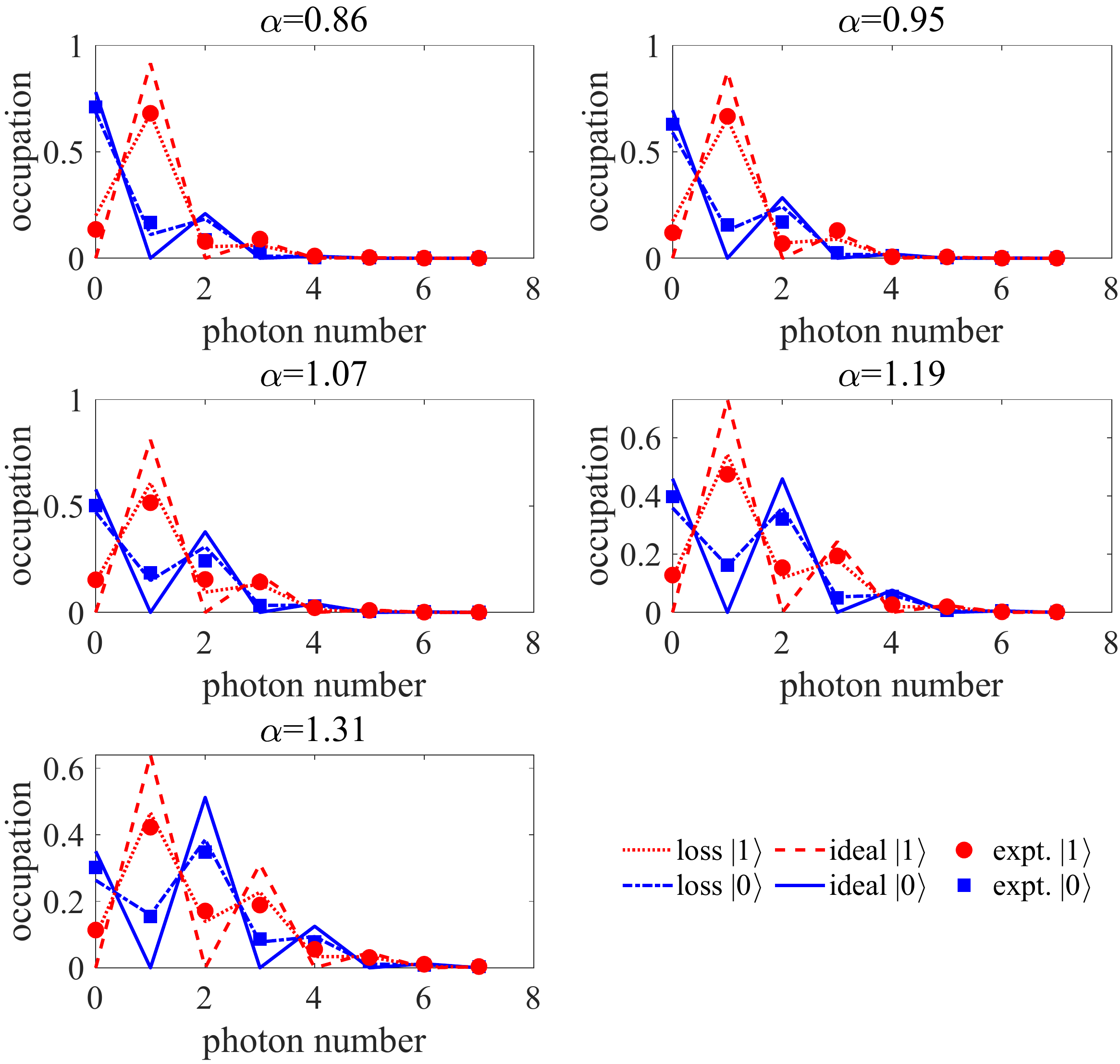}
\caption{\textbf{Photon number distribution with varied $\alpha$.} The derived photon number distributions among different Fock states for the odd (conditioned on qubit state in $\ket{1}$) and even (conditioned on qubit state in $\ket{0}$) cat states of different sizes. The corresponding $\alpha$ is labeled above each sub-panels. The blue squares and the red circles show the experimental data conditioned on the qubit state in $\ket{0}$ and $\ket{1}$, respectively. The blue solid lines and the red dashed lines are theoretical results based on the corresponding ideal states conditioned on the qubit state in $\ket{0}$ and $\ket{1}$, respectively. The blue dotted dashed lines and the red dotted lines are theoretical results considering possible experimental loss and decoherence conditioned on the qubit state in $\ket{0}$ and $\ket{1}$, respectively.}
\label{distribution_alpha}
\end{figure}

\begin{figure}[!tbp]
\centering
\includegraphics[width=1\linewidth]{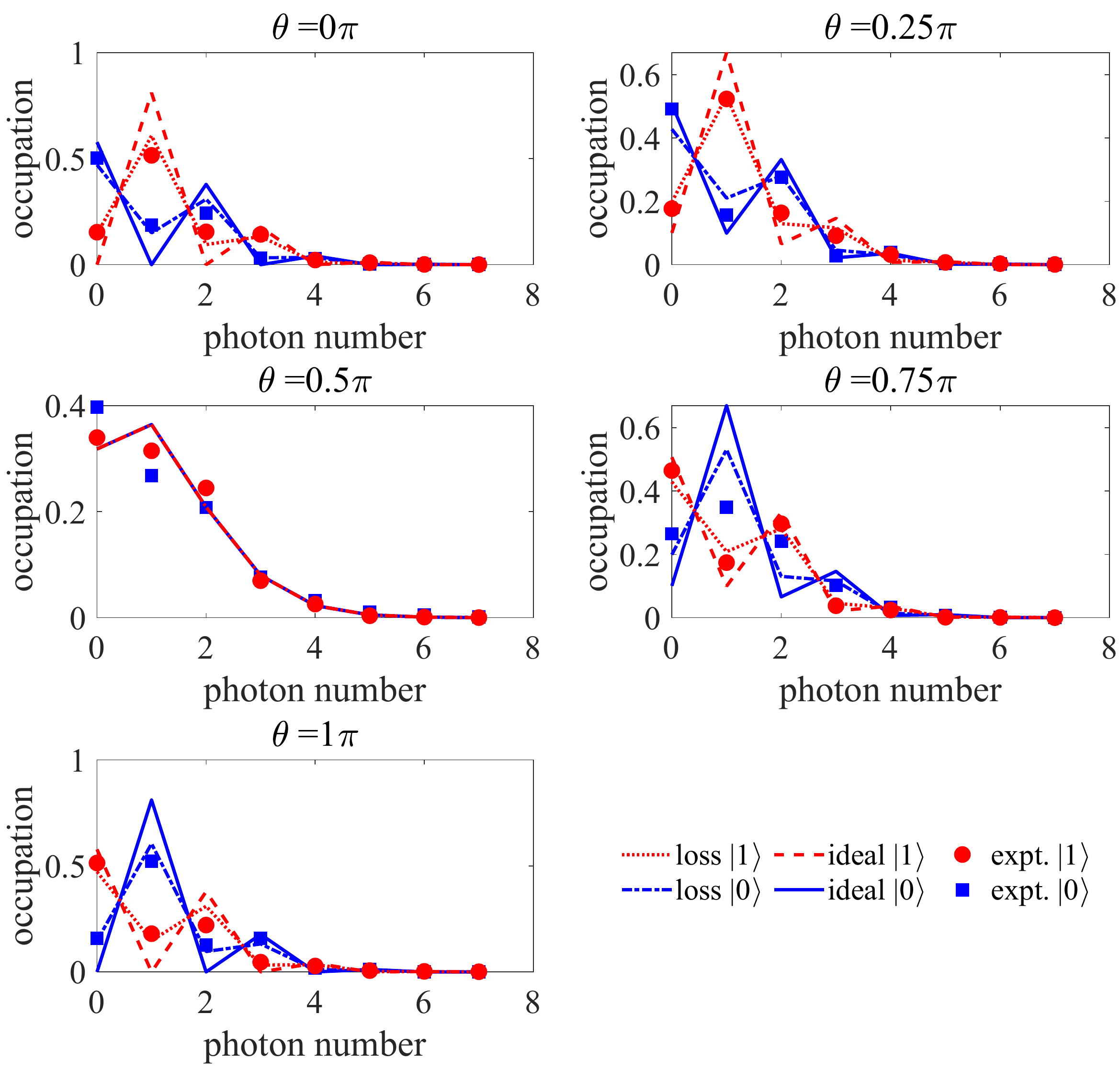}
\caption{\textbf{Photon number distribution with varied $\theta$.} Photon number distributions among different Fock states for superposition states with varied $\theta$ from 0 to $\pi$, and using $\alpha = 1.07$ and $\xi = \pi/2$. The blue squares and the red circles show the experimental data conditioned on the qubit state in $\ket{0}$ and $\ket{1}$, respectively. The blue solid lines and the red dashed lines are theoretical results based on the corresponding ideal states conditioned on the qubit state in $\ket{0}$ and $\ket{1}$, respectively. The blue dotted dashed lines and the red dotted lines are theoretical results considering possible experimental loss and decoherence conditioned on the qubit state in $\ket{0}$ and $\ket{1}$, respectively.}
\label{distribution_theta}
\end{figure}

\begin{figure}[!tbp]
\centering
\includegraphics[width=1\linewidth]{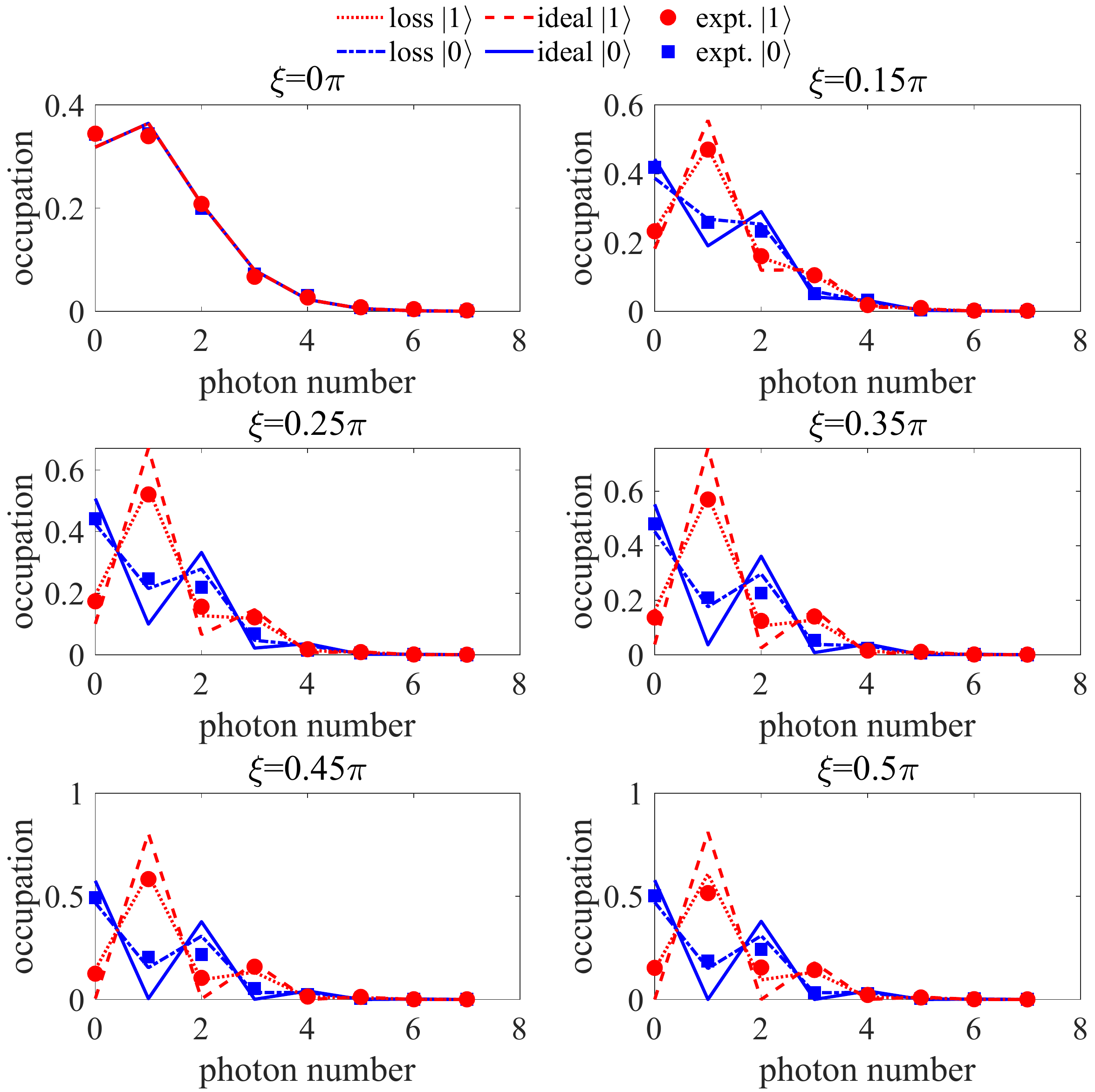}
\caption{\textbf{Photon number distribution with varied $\xi$.} Photon number distributions among different Fock states for superposition states with varied $\xi$ from 0 to $\pi/2$, and using $\alpha = 1.07$ and $\theta = 0$. The blue squares and the red circles show the experimental data conditioned on the qubit state in $\ket{0}$ and $\ket{1}$, respectively. The blue solid lines and the red dashed lines are theoretical results based on the corresponding ideal states conditioned on the qubit state in $\ket{0}$ and $\ket{1}$, respectively. The blue dotted dashed lines and the red dotted lines are theoretical results considering possible experimental loss and decoherence conditioned on the qubit state in $\ket{0}$ and $\ket{1}$, respectively.}
\label{distribution_xi}
\end{figure}

The photon number distributions can be extracted from the diagonal terms of the reconstructed density matrices, which are plotted in Fig.~\ref{distribution_alpha}, Fig.~\ref{distribution_theta} and Fig.~\ref{distribution_xi}.

Fig.~\ref{QQQ} shows the Mandel's Q parameters for the prepared states with different sets of ($\alpha$, $\theta$, $\xi$), conditioned on the qubit state either in $\ket{0}$ or in $\ket{1}$. When using $\theta=0$ and $\xi=\pi/2$, we would have the even and odd cat states conditioned on the qubit state in $\ket{0}$ and $\ket{1}$, respectively. In Fig.~\ref{QQQ}(a), we show the calculated Mandel's Q for the even and odd states with varied sizes. We could see the even cat states show super-Poissonian distribution while the odd states are of sub-Poissonian distribution, which agrees well with the theory. With an increasing $\alpha$, the photon statistics of both even cat states and odd cat states evolve to a Poissonian distribution, which is due to the stronger dephasing effect of the cat states with larger size. 
In Fig.~\ref{distribution_alpha}, we show the photon number distribution among the Fock states with varied $\alpha$. One could see that the even (odd) Fock states are preferably occupied for the prepared cat states. With an increasing $\alpha$, such a selective feature of photon number occupation tends to get smaller.

In Fig.~\ref{QQQ}(b), we show the Mandel's Q for the superposition states with different $\theta$, while using $\xi=\pi/2$ and $\alpha=1.07$. Conditioned on the qubit states in $\ket{0}$, the prepared states evolve from an even cat state to an odd cat state, and vice versa for qubit state in $\ket{1}$. For both cases we could see a continuous transition between a super-Poissonian distribution and a sub-Poissonian distribution for the photon statistics of the prepared states. In Fig.~\ref{distribution_theta}, we show the photon number occupation for the prepared states with different $\theta$. One could see that for the quantum states conditioned on the qubit state in $\ket{0}$ ($\ket{1}$), the photon fields change from an even-Fock-state (odd-Fock-state) occupation to an odd-Fock-state (even-Fock-state) occupation. For $\theta = \pi/2$, the photon fields show little preference of even or odd Fock state occupation, as expected for YS states.

In Fig.~\ref{QQQ}(c), the Mandel's Q for the superposition states with varied $\xi$ when using  $\theta=0$ and $\alpha=1.07$ are presented. One could see the photon field statistic evolves from a Poissonian distribution to either a super-Poissonian distribution conditioned on the qubit state in $\ket{0}$, or sub-Poissonian distribution conditioned on the qubit state in $\ket{1}$, which indicates the prepared state changes from a coherent state to either an even cat state or an odd cat state with an increasing $\xi$. A more direct evidence of such an transition can be found in Fig.~\ref{distribution_xi}, where a Poissonian distribution with no preference of even or odd Fock state occupations gradually evolves to an alternate distribution among the even or odd Fock states conditioned on the qubit state in $\ket{0}$ or $\ket{1}$, respectively.

\clearpage
\pagebreak


%

\end{document}